\documentclass[prd,
,aps,amsmath,amssymb,nofootinbib,preprintnumbers,
]
{revtex4}

\usepackage{graphicx}
\usepackage{dcolumn}
\usepackage{bm}
\usepackage{amsmath}
\usepackage{amsfonts}
\usepackage{subfigure}

\def\ls{\mathrel{\lower4pt\vbox{\lineskip=0pt\baselineskip=0pt
           \hbox{$<$}\hbox{$\sim$}}}}
\def\gs{\mathrel{\lower4pt\vbox{\lineskip=0pt\baselineskip=0pt
           \hbox{$>$}\hbox{$\sim$}}}}
\def\drawbox#1#2{\hrule height#2pt
\hbox{\vrule width#2pt height#1pt \kern#1pt
              \vrule width#2pt}
              \hrule height#2pt}

\def\Asym#1#2{\vcenter{\vbox{\drawbox{#1}{#2}
              \kern-#2pt       
              \drawbox{#1}{#2}}}}

\newcommand{\beq}{\begin{equation}}
\newcommand{\eeq}{\end{equation}}
\newcommand{\dm}{\partial_\mu}
\newcommand{\nn}{\nonumber \\}

\newcommand{\hI}{\hspace{1cm}}

\newcommand{\hV}{\hspace{.5cm}}

\begin{document}
\vspace*{-5mm}
\begin{flushright}
preprint IFT-UAM/CSIC-08-93
\end{flushright}
\vspace*{5mm}
\title{Preheating in the Standard Model with the Higgs-Inflaton coupled to gravity}
\author{
 Juan Garc\'{\i}a-Bellido,\footnote{E-mail: juan.garciabellido@uam.es}\hspace{1mm} 
  Daniel G. Figueroa\footnote{E-mail: daniel.figueroa@uam.es} \hspace{0.5mm} and
  \hspace{0.5mm} Javier Rubio\footnote{E-mail: javier.rubio@uam.es} 
  }
\affiliation{
  Instituto de F\'{\i}sica Te\'orica CSIC-UAM, Universidad Aut\'onoma de Madrid, 
 Cantoblanco 28049 Madrid, Spain
 }
\begin{abstract}
We study the details of preheating in an inflationary scenario in which the Standard Model Higgs, strongly non-minimally coupled to gravity, plays the role of the inflaton. We find that the Universe does not reheat immediately through perturbative decays, but rather initiate a complex process in which perturbative and non-perturbative effects are mixed. The Higgs condensate starts oscillating around the minimum of its potential, producing $W$ and $Z$ gauge bosons non-perturbatively, due to violation of the so called adiabaticity condition. However, during each semi-oscillation, the created gauge bosons partially decay (perturbatively) into fermions. The decay of the gauge bosons prevents the development of parametric resonance, since bosons cannot accumulate significantly at the beginning. However, the energy transferred to the decay products of the bosons is not enough to reheat the Universe, so after about a hundred oscillations, the resonance effects will eventually dominate over the perturbative decays. Around the same time (or slightly earlier), backreaction from the gauge bosons into the Higgs condensate will also start to be significant. Soon afterwards, the Universe is filled with the remnant condensate of the Higgs and a non-thermal distribution of fermions and bosons (those of the SM), which redshift as radiation and matter, respectively.
We compute the distribution of the energy budget among all the species present at the time of backreaction. From there on until thermalization, the evolution of the system is highly non-linear and non-perturbative, and will require a careful study via numerical simulations.
\end{abstract}
\maketitle
\vspace*{-2mm}
\section{Introduction}\label{introduction}
Inflation is nowadays a well established paradigm, consistent with all the observations, that solves most of the puzzles of the Hot Big Bang Model in a very simple and elegant way. It is able to explain not only the homogeneity and isotropy of the present Universe on large scales, but also the generation of almost scale invariant primordial perturbations that give rise to the structure formation \cite{inflation}. However, the naturalness of inflation is directly related to the origin of the inflaton. Most of the inflationary models proposed so far require the introduction of new degrees of freedom  to drive inflation. The nature of the inflaton is completely unknown, and its role could be played by any candidate able to imitate a scalar condensate (tipically in the slow-roll regime), such as a fundamental scalar field, a fermionic or vector condensate, or even higher order terms of the curvature invariants. The number of particle physics motivated candidates is as big as the number of extensions of the Standard Model (Grand Unified Theories, supersymmetry, extra dimensions, etc.),  where it is not very difficult to find a field that could play the role of the inflaton~\cite{Lindebook}.

In addition, given a model we must find a graceful exit to inflation and a mechanism to bring the Universe from a cold and empty post-inflationary state to the highly entropic and thermal Friedmann Universe~\cite{reheating}. Unfortunately, the theory of reheating is also far from being complete, since not only the details, but even the overall picture, depend crucially on the different microphysical models. It seems difficult to study the details of reheating in each concrete model without the experimental knowledge of the strength of the interactions among the inflaton and the matter fields. Because of this, most of the work until now has focussed on models encoding the different mechanisms that could play a role in the process, with the strength of the couplings set essentially by hand. 
The relative importance of each one of these mechanisms can only be clarified in light of an underlying particle physics model, able to provide us with the couplings among the inflaton and matter fields. From this point of view it is very difficult to single out a given model of inflation, and even more difficult to understand the details of the reheating process via the experimental access to the couplings. 

We may be very far away from understanding the microphysical mechanisms responsible for inflation, but maybe the natural candidate for being the inflaton was already there long ago. If we do not want to introduce new dynamical degrees of freedom in the theory, apart from those present in the Standard Model (SM) of particle physics, and at the same time we require Lorentz and gauge invariance, we are left with just one possibility: the Higgs field. Early models of inflation in terms of a Higgs-like scalar field $h$ with a  quartic self-interaction potential $\frac{\lambda}{4}h^4$  need an extremely small coupling constant $\lambda\sim 10^{-13}$ \cite{Lindebook}, and are also nowadays excluded at around $3\sigma$ by the present observational data \cite{WMAPquarticinflation}. However, the SM-based inflation may be rescued from these difficulties replacing the usual  Einstein-Hilbert action by a non-minimal coupling of the Higgs field to the Ricci scalar \cite{Mi77,Ad82,Smo79,Ze79,SalopekBardeenBond}, in a scalar-tensor theory fashion 
\begin{equation}
\label{IG}
S_{IG} =\int d^4x \sqrt{-g} \; \xi H^\dagger H R\;.
\end{equation}
The induced gravity ($IG$) action is indeed the most natural generalization of the Standard Model in a curved spacetime, given the relevance of the non-minimal coupling to gravity for renormalizing the theory  \cite{Birrell:1982ix}.  It belongs to a group of theories known as scalar-tensor theories, originally introduced by Brans and Dicke \cite{BrDi61}  to explain the origin of the masses. In those theories  not only the active masses but also the gravitational constant $G$ ,  are determined by the distribution of matter and energy throughout the Universe, in a clear connection with the Mach principle. The interaction that gives rise to the masses should be the gravitational one, since gravity couples to all particles, {\it i.e.} to their masses or energies. The gravitational constant $G$ in these theories is replaced by a scalar function that permeates the whole space-time and interacts with all the ordinary matter content, determining how the later moves through space and time. Any measurements of an object's mass depends therefore on the local value of this new field. 

The successful Higgs mechanism lies precisely in the same direction of the original Mach's idea of producing mass by a gravitational-like  interaction. The  role of the Higgs field in the Standard Model is basically to provide the inertial mass of all matter fields through the local Spontaneous Symmetry Breaking mechanism\footnote{We are not considering here Majorana masses.}. The Higgs boson couples to all the particles in the Standard model in a very specific way, with a strengh proportional to their masses \cite{SModelref}, and mediates a scalar gravitational interaction of Yukawa type~\cite{DeFrGh90,DeFr91}, between those particles which become massive as a consequence of the local Spontaneous Symmetry Breaking.  According to the Equivalence Principle, it seems natural to identify the gravitational and particle physics approaches to the origin of the masses. From this point of view, the induced gravity action (\ref{IG}) would be an indication of a connection between the Higgs, gravity and inertia. Indeed, the action (\ref{IG}) is, at least at the classical level, just a different representation of the Starobinsky's model of inflation \cite{inflation, MuhkanovR2}, where inflation is entirely a property of the gravitational sector. Both representations of the same theory are simply related by a Legendre transformation. This fact, together with the possibility of having an inflationary expansion of the Universe,  makes the model extremely appealing. Unfortunately, the induced gravity model cannot be accepted as a completely satisfactory inflationary scenario, since the gauge bosons acquire a constant mass in the Einstein frame and totally decouple from the Higgs-inflaton field \cite{SalopekBardeenBond}, which translates into an inefficient reheating of the Universe.
Notice however that the action (\ref{IG}) is not the most general one that can be written in a nontrivial background. As was shown in \cite{SalopekBardeenBond,Shaposhnikov:2007} the simultaneous existence of a reduced \textit{bare} Planck mass $M_P$ and a non-minimal coupling of a Symmetry Breaking field to the scalar curvature 
\begin{equation}\label{SHG}
    S_{HG} \equiv \int d^4x \sqrt{-g} \Bigg\{
     \frac{M_P^2}{2}R+\xi H^\dagger H R \Bigg\}\;,
\end{equation}
avoid the decoupling of the gauge bosons and can give rise to an inflationary expansion of the Universe together with a potentially successful reheating. 

In this paper we initiate the study of the reheating process in the model presented in Ref.~\cite{Shaposhnikov:2007}, in which the Symmetry Breaking field is the Standard Model Higgs, strongly non-minimally coupled to gravity and playing the role of the inflaton. The novelty and great advantage of this model is its connection with a well-known microphysical mechanism, hopefully accessible in the near future accelerator experiments. The measurement of the Higgs mass will complete the list of the couplings of the Standard Model and, therefore, one should be able to study all the details of the reheating mechanism. This makes the model under consideration extremely interesting and, potentially, predictive. Reheating in the context of scalar-tensor theories has been studied, in the Hartree approximation by Ref.~\cite{MaedaTorii2000}, and perturbatively by Ref.~\cite{WatanabeKomatsu2007}, but without the Higgs boson playing the role of the inflaton, and therefore without an explicit coupling of the fundamental scalar field to matter.

Studying the details of reheating in this Higgs-Inflaton scenario, we have put special attention to the relative impact of the different mechanisms that can take place. We have found that the Universe does not reheat immediately through perturbative decays, but rather initiate a complex procces in which perturbative and non-perturbative effects are mixed. The Higgs condensate starts oscillating around the minimum of its potential, producing $Z$ and $W$ gauge bosons due to violation of the so called adiabaticity condition. During each semi-oscillation, the non-perturbatively created gauge bosons decay (perturbatively) into fermions. This decay prevents the development of the usual parametric resonance, since bosons do not accumulate significantly at the beginning. The energy transferred to the decay products of the bosons is not enough to reheat the Universe within a few oscillations, and therefore the resonance effects will eventually dominate over the perturbative decays. Around the same time, the backreaction from the gauge bosons into the Higgs condensate will also start to be significant. Soon afterwards, the Universe is filled with the remnant condensate of the Higgs and a non-thermal distribution of fermions and bosons (those of the SM), which redshift as radiation and matter, respectively. 
We end the paper computing the distribution of the energy budget among all the species present at the time of backreaction. From there on until thermalization, the evolution of the system is highly non-linear and non-perturbative, and will require a careful study via numerical simulations, to be described in a future publication.

The paper is organized as follows. In section~\ref{Higgssector} we present the model with the Higgs field non-minimally coupled to the scalar curvature, transform it into a new frame where the action takes the usual Einstein-Hilbert form, and derive an approximate inflationary potential. In section~\ref{mattersector} we study the effect of the conformal transformation in the matter sector, including the interaction among the Higgs, vector bosons and fermions. Section~\ref{reheating} is devoted to the analysis of the different reheating mechanisms, both perturbative and non-perturbative, that can take place, leaving for Section~\ref{stimulateddecays} the analysis of the combined effect of parametric resonance and perturbative decays. We then study the backreaction of the produced particles on the Higgs oscillations and the end of preheating in Section~\ref{backreaction}. Finally the conclusions are presented in Section \ref{conclusions}.

\section{The Standard Model Higgs as the inflaton}\label{Higgssector}

The Glashow-Weinberg-Salam \cite{SModelref} action is divided into four parts: a fermion sector $(F)$ which includes the kinetic terms for the fermions and their interaction with the gauge bosons, a gauge sector $(G)$, including the kinetic terms for the intermediate bosons as well as the gauge fixing and Faddeev-Popov terms, a Spontaneous Symmetry Breaking  sector $(SSB)$, with a Higgs potential and the kinetic term for the Higgs field including its interaction with the gauge fields, and finally, a Yukawa sector $(Y)$, with the interaction among the Higgs and the fermions of the Standard Model,
\begin{equation}
\label{SM}
S_{SM}=  S_{F}+S_{G}+S_{SSB}+S_{Y}\,.
\end{equation}
The simplest versions of this Lagrangian in curved spacetime follow the principles of general covariance and locality for both matter and gravitational sectors. To preserve the fundamental features of the original theory in flat space-time, one must also require the gauge invariance and other symmetries in flat space-time to hold for the curved space-time theory. The number of possible terms in  the action is unbounded  even in this case and some additional restrictions are needed. A natural requirement could be renormalizability and simplicity. Following this three principles (locality, covariance and restricted dimension), and the previously motivated requirement of not introducing new dynamical degrees of freedom, the form of the action is fixed, except for the values of some new parameters to be determined by the physics.
This procedure leads to the non-minimal Lagrangian for the Standard Model in the presence of gravity, given by 
\begin{equation}
\label{SMG}
  S_{\mathrm{SMG}}= S_{SM}+S_{HG} \;,
\end{equation}
where $S_{\mathrm{SM}}$ is the Standard Model part~(\ref{SM}) defined above, and $S_{HG}$ is the new Higgs-gravity sector, given by~ Eq. (\ref{SHG}). Here $M_P = (8\pi\mathrm{G})^{-1/2}$ is the reduced Planck mass,
$R$ the Ricci scalar, $H$ the Higgs field, and $\xi$ is the announced non-minimal coupling
constant. As showed in Ref.~\cite{Shaposhnikov:2007}, the parameters $\xi$ and the self-coupling $\lambda$ of the Higgs potential are related by $\xi\simeq 49000\sqrt{\lambda}$. 
In the unitary gauge, $H=h/\sqrt{2}$, and neglecting all gauge interactions for the time being, the Lagrangian for the Higgs-gravity sector in the so-called Jordan (J) frame takes the form
\begin{equation}\label{higgslagrangJ}
 S_{HG}+S_{SSB}  \supset\int d^4x \sqrt{-g} \Big[f(h)R - 
 \frac{1}{2}g^{\mu\nu}\partial_\mu \,h\partial_\nu h - U(h) \Big] \,, 
\end{equation}
where $f(h)=(M_P^2+\xi h^2)/2$, and 
\begin{equation}\label{potentialJ}
U(h)=\frac{\lambda}{4}\left(h^2-v^2\right)^2\,,
\end{equation}
is the usual Higgs potential of the Standard Model, with vev $v=246$ GeV.

In order to get rid of the non-minimal coupling to gravity, we proceed as usual, performing a conformal transformation \cite{Maeda1989}
\begin{equation} \label{ct}
g_{\mu\nu} \rightarrow \tilde{g}_{\mu\nu}=\Omega^2 g_{\mu\nu} \; ,
\end{equation}
such that we obtain the Lagrangian in the so-called Einstein (E) frame 
\begin{equation}
\label{higgsLagrangianE}
 S^E_{HG}+S^E_{SBS}\supset\int d^4x \sqrt{-\tilde{g}} \, \Bigg\{
     \frac{f(h)}{\Omega^2}\left[\tilde{R}+3\tilde{g}^{\mu\nu}\tilde{\nabla}_\mu \tilde{\nabla}_\nu \ln\Omega^2-\frac{3}{2}\tilde{g}^{\mu\nu}\tilde{\nabla}_\mu \ln \Omega^2 \tilde{\nabla}_\nu \ln \Omega^2\right]
   -\frac{\tilde{\dm} h\tilde{\partial}^\mu h}{2\Omega^2}-\frac{1}{\Omega^4}U(h) \Bigg\}.
\end{equation}
The usual Einstein-Hilbert term can then be obtained imposing 
$f(h)/\Omega^2\equiv M_P^2/2$,
which implies the following relation between the conformal transformation and the Higgs field 
\begin{equation}
\label{relwithomega} 
 \Omega^2(h) = 1 + \frac{\xi h^2}{M_P^2}\,.
\end{equation}
This allows us to write the Lagrangian (\ref{higgsLagrangianE}) completely in terms of $h$
\begin{equation}
\label{higgsLagrangianE2}
 S^E_{HG}+S^E_{SSB} \supset \int d^4x \sqrt{-\tilde{g}} \, \Bigg\{   \frac{M_P^2}{2}\tilde{R}
    -\frac{1}{2}\Big[ \frac{\Omega^2+6\xi^2h^2/M_P^2}{\Omega^4}\Big]
    \tilde g^{\mu\nu}\partial_\mu h\, \partial_\nu h - {1\over \Omega^4}U(h)\Bigg\}, 
\end{equation}
where we have neglected a total derivative that does not contribute to the equations of motion. 
As we will be working in the Einstein frame from now on, we will skip over the tilde in all the variables to simplify the notation.

Notice that the conformal transformation (\ref{ct}) leads to a non-minimal kinetic term for the Higgs
field, which can be reduced to a canonical one by making the transformation
\begin{equation} \label{relchih}
  \frac{d\chi}{dh}=\sqrt{\frac{\Omega^2+6\xi^2h^2/M_P^2}{\Omega^4}}=
  \sqrt{{1+\xi(1+6\xi)h^2/M_P^2 \over (1+\xi h^2/M_P^2)^2}}\,,
\end{equation}
where $\chi$ is a new scalar field. Doing this, the total action in the Einstein frame, without taking into account the gauge interactions, is simply
\begin{equation}
  \label{higgsLagrangianE3}
    S^E_{HG}+S^E_{SSB} \supset \int d^4x\sqrt{- {g}} \, \Bigg[ \frac{M_P^2}{2} {R}
    - {1\over2}  g^{\mu\nu}\partial_\mu \chi\,\partial_\nu \chi - V(\chi)  \Bigg] \,,
\end{equation}
with 
\begin{equation}
\label{potentialE}
V(\chi)\equiv\frac{1}{\Omega^4(\chi)}U(h(\chi))\;, 
\end{equation}
the potential in terms of the new field $\chi$. To find the explicit form of the potential in this new variable $\chi$, we must find the expresion of $h$ in terms of $\chi$. This can be done by integrating Eq. (\ref{relchih}), whose general solution is given by
\begin{equation}\label{sol}
\frac{\sqrt{\xi}}{M_P}\chi(h)=\sqrt{1+6\xi}\sinh^{-1}\left(\sqrt{1+6\xi}u\right)-\sqrt{6\xi}\sinh^{-1}\left(\sqrt{6\xi}\frac{u}{\sqrt{1+u^2}}\right)\;,
\end{equation}
where $u\equiv\sqrt{\xi} h/M_P$. Since $\xi\gg 1$, we can take $1+6\xi \approx 6\xi$ and, using the identity
$\sinh^{-1}x=\ln(x+\sqrt{x^2+1})$ for $ -\infty<x<\infty $, we can approximate Eq. (\ref{sol}) by
\begin{equation}\label{approxsol}
 \frac{\sqrt{\xi}}{M_P}\chi(h) \approx \sqrt{6\xi}\ln (1+u^2)^{1/2}\;,
\end{equation}
or, equivalently,
\begin{equation}\label{relchih2}
 \Omega^2=e^{\alpha\kappa\chi}\;,
\end{equation}
where $\alpha=\sqrt{2/3}$ and $\kappa=M_P^{-1}$. The $\chi$ field is therefore directly related in this aproximation (just in the limit $\xi \gg 1$ and far from $u=0$) to the conformal transformation $\Omega$ in a very simple way  and the inflationary potential (\ref{potentialE}) is just given by
\begin{equation}\label{potentialE1}
 V(\chi)=\Omega^{-4}U(h)=\frac{\lambda M_P^4}{4\xi^2}\left[e^{\alpha\kappa\chi}-\left(1+\xi\frac{v^2}{M_P^2}\right)\right]^2e^{-2\alpha\kappa\chi}\;.
\end{equation}
Since $v \ll M_p$, then $1+\xi\frac{v^2}{M_P^2}\approx 1$ and we can savely ignore the \textit{vev} for the evolution during inflation and preheating, and simply consider the potential 
\begin{equation}
\label{potentialE2}
 V(\chi)=\frac{\lambda M_P^4}{4\xi^2}\Big(1-e^{-\alpha\kappa\chi}\Big)^2\;.
\end{equation}
\begin{figure}
\centering
\includegraphics[width=12cm]{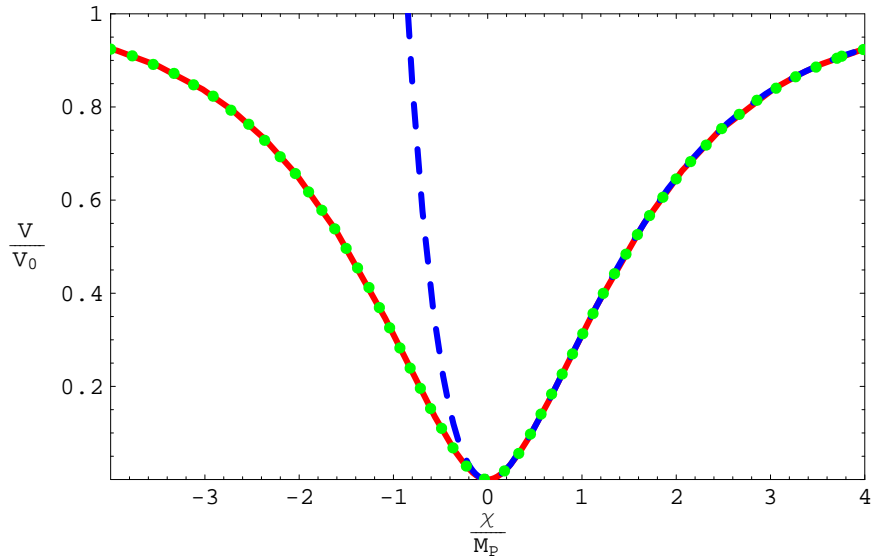}
\caption{Comparative plot of the exact solution (red continuous line) obtained parametrically from Eq.(\ref{sol}), the analytic formula (\ref{potentialE2}) for the potential (blue dashed line), and their parametrization (\ref{potentialE3}) (green dotted line).}
 \label{comparison}
\end{figure}
Notice that the previous potential only parametrizes partially the original potential (\ref{potentialJ}), since it neglects the region $\chi<0$, as can be seen in Fig.\ref{comparison}, where we compare the exact solution (red continuous line) obtained parametrically from Eq. (\ref{sol}), with the analitic formula (\ref{potentialE2}) (blue dashed line). Both solutions agree very well in the region of positive $\chi$ but differ substancially for $\chi<0$. The conformal transformation is even ill-defined in the negative field region. From Eq. (\ref{relwithomega}) and Eq. (\ref{relchih2}) we have 
\begin{equation}
\frac{\xi h^2}{M_P^2}=\Omega^2-1=e^{\alpha\kappa\chi}-1=(1-e^{-\alpha\kappa\chi}) e^{\alpha\kappa\chi}\;,
\end{equation}
which is inconsistent, since the left-hand side of this equation is positive definite, while the right hand is negative definite for $\chi<0$. Taking this into account, 
in order to study the different mechanisms of the 
post-inflationary regime, 
we will then use the parametrization
\begin{equation}
\label{potentialE3}
 V(\chi)=\frac{\lambda M_P^4}{4\xi^2}\left(1-e^{-\alpha\kappa\vert\chi\vert}\right)^2\;.
\end{equation}
which correctly describes the potential obtained from Eq.~(\ref{sol}), for the whole field range of interest. In Fig.~\ref{comparison}, this parametrization (green dotted line) is again compared to the exact solution (red continuous line) obtained from Eq. (\ref{sol}). 
Around then minimum, the potential (\ref{potentialE3}) can be approximated as \begin{eqnarray}\label{potentialEapprox}
V(\chi)= {1\over2}M^2\chi^2 + \Delta V(\chi)\,,
\end{eqnarray} 
where $M^2=\lambda M^2_P/3\xi^2$ is the typical frequency of oscillation and $\Delta V$ are some corrections to the quadratic approximation, which soon become negligible after inflation ends, see  section~\ref{reheating}. Note, nevertheless, that approximations~(\ref{approxsol}),(\ref{relchih2}) and therefore parametrization~(\ref{potentialE3}), do not describe correctly the potential for very small values of the field $v \ll  \chi \ll M_P/\xi$. As can be read from Eq.~(\ref{relchih}), for $\vert\chi\vert \ll \chi_t \equiv M_P/\xi$, we have $\frac{d\chi}{dh} \approx 1$ and therefore, there is a transition in the potential 
from~(\ref{potentialEapprox}) to $V(\chi) \approx \frac{\lambda}{4}\chi^4$. However, as will be shown in section \ref{parametricreheating}, the transition region, $|\chi| < \chi_t$, is several orders of magnitude smaller than the non-adiabaticity region, $|\chi| < \chi_a$, see Eq.~(\ref{chia}), inside which the concept of particle is not properly defined. Therefore, from now on we will neglect the change in the behaviour of the potential (from $\frac{1}{2}M^2\chi^2$ to $\frac{\lambda}{4}\chi^4$) in this ``small'' field region, since $\chi_t \ll \chi_a$. See section~\ref{reheating} for more details.

The analysis of the inflationary potential (\ref{potentialE3}) can be performed either in the Jordan~\cite{Makino1991,Fakir1992} or in the Einstein frame~\cite{Shaposhnikov:2007} with the same result. The slow roll parameters can be expressed analytically, in the limit of $h^2\gg M_P^2/\xi\gg v^2$, as a function of $\chi$,
\begin{equation}\label{slowrollparam}
  \epsilon  =  \frac{M_P^2}{2}\left(\frac{V'(\chi)}{V(\chi)}\right)^2
  =\frac{2\alpha^2}{(e^{\alpha\kappa\chi}-1)^2}\;,
   \hspace{15mm}
  \eta  =  M_P^2\frac{V''(\chi)}{V(\chi)}
  =\frac{2\alpha^2(2-e^{\alpha\kappa\chi})}{(e^{\alpha\kappa\chi}-1)^2} \,.
\end{equation}
The slow roll regime of inflation will end when $\epsilon \simeq 1$, which correspond to the field value 
\begin{equation}\label{chiend}
\chi_{\rm end}=\frac{1}{\alpha\kappa}\ln\left(1+\frac{2}{\sqrt{3}}\right)\,.
\end{equation}
Note that the slow roll parameter $\eta$ is then negative, $\eta_{\rm end}=1-\frac{2}{\sqrt{3}}<0$, so there is a small region of negative (mass squared)
curvature in the potential just after the end of inflation. The effective curvature of the potential will be negative until $ \chi_*=\frac{1}{\alpha\kappa}\ln 2$, which corresponds to the inflection point, given by $\eta_*=0$. 

During the slow-roll regime $H_{SR}=\frac{\kappa}{\sqrt{3}}V^{1/2}$, which evaluated at $N=60$ e-folds is approximately given by $H_{60}\simeq\frac{M}{2}$, where $M$ defines the natural inflationary energy scale of this model as well as the frequency of oscillations~(\ref{potentialEapprox}) during reheating. At the end of inflation $H_{\rm end}=\frac{\kappa}{\sqrt{2}}V^{1/2}$ or, equivalently, $H_{\rm end}\simeq\frac{2}{3}H_{60}=\frac{M}{3}$. 

The radiative corrections for a model containing a single scalar field $h$ non-minimally coupled to gravity were generically calculated in Ref.~\cite{BK}. The specific radiative corrections for the model under consideration, estimated in Ref.~\cite{Shaposhnikov:2007}, were recently reviewed in Ref.~\cite{BKS2008}. In what follows until the end of this section, we will summarize the results of this last work. For large $\xi$ and slow varying $h$ the main contribution comes from loops of the matter \cite{BK,BKK} and the effective action can be calculated by a local $1/m^2$-expansion in powers of the curvature and its gradients and the gradients of the Higgs field, obtaining \cite{BK,BKS2008}
\begin{equation}
S[g_{\mu\nu},h]=\int d^{4}x\,g^{1/2}\left(-U(h)+F(h)\,R(g_{\mu\nu}) - \frac12\,G(h)\,(\nabla h)^2\right)\,,\label{effectiveAction}
\end{equation}
where the functions $V(h)$, $U(h)$ and $G(h)$ are given by
\begin{eqnarray}
&&U(h)=\frac\lambda{4}(h^2- v^2)^2 + \frac{\lambda h^4}{128\pi^2}\left(A\ln\frac{h^2}{Q^2} + B\right),\label{potentialCorregido}\\
&&F(h)= \frac12(M_P^2+\xi h^{2}) + \frac{h^2}{384\pi^2}\left(C\ln\frac{h^2}{Q^2} + D\right), \label{gravityCorrected}\\
&&G(h)=1+\frac{1}{192\pi^2}\left(F\ln\frac{h^2}{Q^2} + E \right)\;.\label{kineticCorrected}
\end{eqnarray}
Here $A, B, C, D, E$ and $F$ are different combinations of the Higgs, gauge and Yukawa couplings and their logarithms \cite{BK,Shore,AlStar1}, and $Q$ is the normalization scale. Following Ref.~\cite{BKS2008}, for the analysis of inflation we will just consider 
the explict form of the combination $A$,
\begin{eqnarray}\label{scalingA}
A = \frac{2}{\lambda}
\left(3 \sum_{A} g_{A}^{4} - 
\sum_{f} y_f^{4}\right)\;, \label{Aconstant}
\end{eqnarray}
which is related with the local conformal anomaly. The factor $3$ accounts for the polarizations of the gauge bosons and a similar factor $4$ for the fermions has been taken into account.

The new inflationary potential in the Einstein frame for a Higgs field with non-minimal coupling $\xi\gg1$ and mean value much greater than the minimum of the
classical potential, is given by \cite{BKS2008}
\begin{eqnarray}
\tilde{V}(\tilde\chi)=\left.\left(\frac{M_P^2}{2}\right)^2\frac{U(h)}{F^2(h)}\,\right|_{\,h=h(\tilde\chi)}~. \label{tildeV}
\end{eqnarray}
The corresponding slow roll parameters are
\begin{eqnarray}\label{TildeEpsilon}
\tilde\epsilon = \frac{4M_P^4}{3\xi^2h^4}\,\left(1+\frac{h^2}{h_I^2}\right)^2 = \frac43\left(\frac{M_P^2}{\xi\,h^2}+\frac{A}{64\pi^2}\!\right)^2,\hI  \tilde\eta =-\frac{4M_P^2}{3\xi h^2}~\,,\label{tildeEta}
\end{eqnarray}
with $h_I^2=\frac{64\pi^2 M_P^2}{\xi A}$. Using the WMAP+BAO+SN 
constraint~\cite{WMAPnorm} at the $2\sigma$ confidence level gives a value for the spectral index~\cite{BKS2008}
\begin{equation}
0.934 <n_s(k_0)<0.988~. \label{ns}
\end{equation}
For $N(k_0)=60$ the value for $A$ and for the tensor-to-scalar ratio is \cite{BKS2008}
\begin{eqnarray}
-12.4<A<14.1 ~,  \label{rangoA} \\
0.0006<r<0.015\,, \label{rangor}
\end{eqnarray}
being the spectral scalar index running completely negligible \cite{BKS2008}
\begin{equation}
 -5.6 < \alpha \times 10^4 < -4.3\;.
\end{equation}
Let us compare this window with the one obtained from the Standard Model coupling constants at the scale 
$M$, and obtain the Higgs self-coupling at such scale. This differs from the analysis performed in~\cite{BKS2008}, where the coupling constants were evaluated at the electroweak scale instead of at the characteristic energy scale of inflation, $M$. At that scale the coupling constants for the gauge groups $SU(2)_L$ and $U(1)_Y$ are roughly equal $g_1^2\approx g_2^2 \approx 0.3$ and $\cos^2\theta_W=\sin^2\theta_W \approx 1/\sqrt{2}$. The total anomalous scaling constant at that scale, see Eq.~(\ref{scalingA}), in terms of $g_2$ and $\cos\theta_W$, 
\begin{eqnarray}\label{scalingASM}
A \approx\frac{6}{\lambda}
\left[ \frac{g_{2}^{4}}{8}\left(1+\frac{1}{2}\cos\theta_W^{-4}\right) -  y_f^{4}\right]\;,   \label{ASM}
\end{eqnarray}
together with the bounds obtained from the WMAP+BAO+SN $2\sigma$ c.l. constraints \cite{WMAPnorm}, see Eq.~(\ref{rangoA}), give us the following range for the self-coupling of the Higgs field at the scale $M$,
\begin{equation}
 -0.424 < \lambda(M) < 0.482\;. 
\end{equation}
This range can be propagated back to obtain a value at the scale $M_Z^2$ through the renomalization group equations, \begin{equation}
\lambda^{-1}(\mu)=\lambda^{-1}(M)+\frac{3}{4\pi^2}\log\frac{M^2}{\mu^2}\,.
\end{equation}
Note that we have neglected the effect of the gauge and Yukawa couplings, since the complete solution of the renormalization group equations is out of the scope of this preliminary study. If we integrate this equation and take into account the present observational bounds~\cite{PDG2008}, we obtain a Higgs mass in the range
\begin{equation}\label{HiggsRange}
114.5 \ {\rm GeV} < m_H < 275 \ {\rm GeV}\,,
\end{equation}
safely within the detection range of the Large Hadron Collider at CERN. In the absence of an actual measurement of the Higgs self-coupling $\lambda$, for the analysis of sections~\ref{reheating} and thereafter, we will just take different values compatible with the above range~(\ref{HiggsRange}).

\section{The Standard Model matter sector in the Einstein frame}\label{mattersector}

The length scales are conventionally defined in such a way
that elementary particle masses are the same for all times and in all
places. This implies for instance that if under a conformal transformation the Lagrangian of a free particle transforms as 
\begin{equation}\label{1part}
{\cal L}_{1P}=\int m ds\longrightarrow  \tilde{\cal L}_{1P}=\int \frac{m }{\Omega}\tilde{ds}\,,
\end{equation}
the mass should be accordingly redefined  as $\tilde{m}\equiv \frac{m}{\Omega}$ to express it in the new system of units. The previous argument applies also for classical fields \cite{MaedaFujiibook}. The rescaling of all fields (including the metric tensor) with an arbitrary space-time dependent factor 
$\Omega$, taken with a proper conformal weight for each field, will leave the physics unaffected. The physical interpretation of this symmetry is clear:  It changes all dimensional quantities (lengths, masses, etc.) in every point of the space--time   leaving their ratios unchanged. 

In this section we will apply the previous prescription to the different sectors of the action (\ref{SM}). Consider for instance the Spontaneous Symmetry Breaking sector, responsible of the masses of the intermediate gauge bosons $W$ and $Z$
\begin{equation}\label{SSBJ}
S_{SSB}\supset -\int d^4x \sqrt{-g} \Big\{ m_W^2 W_{\mu}^+ W^{\mu -}+\frac{1}{2} m_Z^2 Z_\mu Z^\mu  \Big\}\,.
\end{equation}
In the Standard Model, the masses of the $SU(2)$ bosons are due to a Spontaneous Symmetry Breaking mechanism, realized by the constant \textit{vev} of the Higgs field, and therefore are constant. In our case the Higgs field evolves with time giving rise to variable effective masses for the gauge bosons
\begin{equation}
  \label{massesWZ}
 m_W=\frac{g_2 h}{2} \;, \hspace{15mm} m_Z=\frac{m_W}{\cos\theta_W}  \;,
\end{equation}
with $\theta_W$ the Weinberg angle defined as $\theta_W=\tan^{-1}(g_1/g_2)$ and $g_1$ and $g_2$ are the coupling constants corresponding to $U(1)_Y$ and $SU(2)_L$ at the scale $M$, where the relevant physical processes during preheating will take place. As mentioned before, numerically this correspond to a value $g^2_1 \approx g^2_2 \approx 0.30$, which implies $\sin^2\theta_W = \cos^2\theta_W \approx 1/\sqrt{2}$. From now on we will use these values for numerical estimations. In agreement with the above prescription for transforming masses and fields, the action (\ref{SSBJ}) preserves its form under the conformal transformation 
\begin{equation}\label{SSBE}
S^E_{SSB}\supset -\int d^4x \sqrt{-\tilde{g}} \Big\{ \tilde{m}_W^2 \tilde{W}_{\mu}^+ \tilde{W}^{\mu -}+\frac{1}{2} \tilde{m}_Z^2 \tilde{Z}_\mu \tilde{Z}^\mu  \Big\}\;,
\end{equation}
providing that we redefine the fields and masses with the corresponding conformal weights as
\begin{equation}
 \tilde{W}_\mu^{\pm}\equiv\frac{{W}_\mu^{\pm}}{\Omega}\; ,  \hspace{10mm} \tilde{Z}_\mu\equiv\frac{{Z}_\mu}{\Omega}
 \;,  \hspace{10mm}  \label{redefmasWZ}
 \tilde{m}^2_W=\frac{m^2_W}{\Omega^2}=\frac{g^2_2 M_P^2(1-e^{-\alpha\kappa\vert\chi\vert})}{4\xi}\;, \hspace{10mm} 
\tilde{m}^2_Z=\frac{\tilde{m}^2_W}{\cos^2\theta_W} 
 \;.
\end{equation}
The same can be applied to the interactions between fermions and gauge bosons. Let us consider, for instance, 
\begin{equation}\label{F}
S_F = S_{NC}+S_{CC} \supset \int d^4x \sqrt{-g} \left\{\frac{g_2}{\sqrt{2}}W_\mu^+J^-_\mu +\frac{g_2}{\sqrt{2}}W_\mu^-J^+_\mu+ \frac{g_2}{\cos\theta_W} Z_\mu J_Z^\mu\right\} \;,
\end{equation}
where $J^-_\mu\equiv \bar{ d}_L\gamma^\mu { u}_L, J^+_\mu\equiv {\bar{ u}}_L\gamma^\mu {{d}}_L\;,$
are the charged currents carrying the information about the couplings of the $W^\pm$ to the Standard Model fermions, and
\begin{equation}\label{NCJ}
J_Z^\mu\equiv  \frac{1}{2}{\bar{u}}_L\gamma^\mu {{u}}_L-\frac{1}{2}{\bar{d}}_L\gamma^\mu {{d}}_L-\frac{2\sin^2\theta_W }{3}{\bar{u}}_L\gamma^\mu {{u}}_L+\frac{\sin^2\theta_W }{3} {\bar{d}}_L\gamma^\mu {{d}}_L \;.
\end{equation} 
is the neutral current with the information of the couplings of the $Z$ boson.
In the Einstein frame the action (\ref{F}) preserves its form
\begin{equation}\label{JE}
S^E_F= S^E_{NC}+S^E_{CC} \supset \int d^4x \sqrt{-\tilde{g}}\left\{\frac{g_2}{\sqrt{2}}\tilde{W}_\mu^+\tilde{J}^-_\mu +\frac{g_2}{\sqrt{2}}\tilde{W}_\mu^-\tilde{J}^+_\mu+\frac{g_2}{\cos\theta_W}  \tilde{Z}_\mu \tilde{J}_Z^\mu\right\} \;, 
\end{equation}
as long as we redefine the currents as
\begin{equation}\label{NCE}
\tilde{J_Z^\mu}\equiv\frac{J_Z^\mu}{\Omega^3} \;, \hspace{15mm} \tilde{J}^\pm_\mu\equiv \frac{J^\pm_\mu}{\Omega^3}\;,
\end{equation}
which is equivalent to redefine the Dirac fields
\begin{equation}
\label{redefspinors2}
 \tilde{{d}}\equiv\frac{{d}}{\Omega^{3/2}}\;, \hspace{15mm}  \tilde{{u}}\equiv\frac{{u}}{\Omega^{3/2}}
 \;.
\end{equation}
Finally, concerning the Yukawa sector, we have, for a given family of the quark sector, 
\begin{equation}\label{yukawaJ}
S_{Y}\supset -\int d^4x \sqrt{-g} \Big\{ m_d {\bar{d}}{{d}} + m_u { \bar{u}}{ u}\Big\}\;,
\end{equation}
where ${d}$ and ${u}$ stand for down- and up-type quarks respectively. The effective masses 
in the Jordan frame, $m_f=\frac{y_f h}{\sqrt{2}}$, 
become 
\begin{equation}
  \label{redefmassfermion}
 \tilde{m}_f\equiv\frac{y_f M_P}{\sqrt{2\xi}}\left(1-e^{-\alpha\kappa\vert\chi\vert}\right)^{1/2}\;,
\end{equation}
in the Einstein frame.
\\

On the other hand, the total decay widths, summing over all the allowed decay channels in the Standard Model of the $W^{\pm}$ and $Z$ bosons into any pair of fermions and over all the polarizations of the gauge bosons, are given respectively by \cite{Cheng:2006book}
\begin{eqnarray}\label{widthWJ}
\Gamma_{W^+}=\Gamma_{W^-}= \frac{3g_2^2 m_W}{16\pi}\,\hV, \hI\hV
\Gamma_{Z} = \frac{g^2_2 m_Z}{8\pi\cos^2\theta_W}{\rm Lips}\,.
\end{eqnarray} 
where Lips denotes the \textit{Lorentz invariant phase-space} factors
\begin{equation}\label{Lorentzphase}
 {\rm Lips}\equiv\frac{7}{4}-\frac{11}{3}\sin^2\theta_W+\frac{49}{9}\sin^4\theta_W\;.
\end{equation}
The decay rates will preserve their functional form in the Einstein frame, being only changed through the conformal transformation of the masses, i.e.
\begin{eqnarray}
\label{widthWE}
\Gamma_{W^\pm}^E = \frac{3g_2^2 \tilde{m}_W}{16\pi} = \frac{3g_2^3M_p}{32\pi\xi^{1/2}}\left(1-e^{-\alpha\kappa|\chi|}\right)^{1/2} = \frac{3\cos^3\theta_W}{2{\rm Lips}}\,\Gamma^E_{Z}
\end{eqnarray} 
where $\tilde{m}_{W^\pm}, \tilde{m}_Z \propto \left(1-e^{-\alpha\kappa|\chi|}\right)^{1\over2}$, are the dynamical masses in the Einstein frame, see Eq.~(\ref{redefmasWZ}).

\section{Reheating in the Standard Model of particle physics}\label{reheating}

In this section we will analize the different mechanisms that could give rise to efficient reheating of the Universe, both perturbative and non-perturbative. The natural mechanism, given the strength of the interactions of the Higgs boson with the Standard Model particles, would be a perturbative reheating process right after the end of slow-roll. However, as we will see,  perturbative reheating is not efficient enough, and non-perturbative effects must be taken into account. Given the shape of the potential~(\ref{potentialE3}), very different (p)reheating mechanisms could in principle take place, from a tachyonic production in the region between the end of inflation and the inflection point~\cite{Tachyonic2001,Felder:2001kt,GarciaBellido:2002}, an instant preheating mechanism~\cite{InstantFKL98} and a parametric resonance effect around the minimum of the potential~\cite{KLS94,KLS97,GBL97}. Therefore, it will crucial to disantangle the contribution of each mechanism and quantify their relative importance.

In order to study the different reheating mechanisms it will be useful to have an approximate expression for the evolution of the inflaton. As mentioned before, we can expand potential~(\ref{potentialE3}) for the range of interest, as 
\begin{eqnarray}\label{potentialEapproxII}
V(\chi) = {1\over2}M^2\chi^2 + \Delta V(\chi)\,,
\end{eqnarray} 
with $M^2=\lambda M^2_P/3\xi^2$ the typical frequency of oscillation. The first terms of the corrections $\Delta V$ to the quadratic potential are given explicitely by
\begin{equation}
\Delta V(\chi)= -\frac{\beta}{3}\vert\chi\vert^3+\frac{\zeta}{4}\chi^4 +{\cal O}(\vert\chi\vert^5)\;,
\end{equation}
with $\beta=\lambda M_P/\sqrt{6}\xi^2$ and $\zeta=7\lambda/27\xi^2$.
The Klein-Gordon equation\index{Klein-Gordon equation} for the inflaton,
\begin{equation}
\ddot\chi + 3H\dot\chi + V'(\chi) = 0\;,
\end{equation}
can then be written, for a power-law evolution ${a\propto t^{p}}$, as
\begin{equation}\label{eqNonLinearHiggs}
t^2\ddot\chi + 3pt\dot\chi + t^2M^2[1+\delta M^2(\chi)]\chi = 0\,,
\end{equation}
where we have neglected the Higgs' interactions with other fields, because they are proportional to the number density of particles of a given species and therefore they will be negligible (see section~\ref{backreaction}) during the first oscillations of the Higgs field. The backreaction of other particles into the dynamics of Higgs will only be relevant once their occupations numbers have grown sufficiently. The non-linear terms of the Higgs' self-interaction, described by
\begin{eqnarray}\label{MassPerturbation}
\delta M^2 \approx -\beta|\chi| + \zeta\chi^2 + {\cal O}(\chi^3)\,,
\end{eqnarray}
will be also negligible from the very beginning of reheating, $|\delta M^2(\chi)| \ll 1$, as we will justify \textit{a posteriori}. Thus, neglecting such a term in the effective equation of $\chi$, the general solution can be expressed as 
\begin{equation}\label{solutionBessel}
\chi(t) = \frac{1}{(Mt)^{\nu}}\left[\,A\,J_{+\nu}(Mt) + B\,J_{-\nu}(Mt)\,\right]\,,
\end{equation}
with $A$ and $B$ constants depending on the initial conditions (end of inflation), and $J_{\pm\nu}(x)$ Bessel functions of order $\pm\nu$, with $\nu = (3p-1)/2$. For a reasonable power index,  $p > {1/3}$ -- for matter $p = 2/3$, while for radiation $p = 1/2$ -- the second term in the right-hand side of Eq.~(\ref{solutionBessel}) diverges in the limit ${Mt \rightarrow 0}$ and therefore should be discarded on physical grounds. The physical solution is then simply given by
\begin{equation}\label{physsolution}
\chi(t) = A\,(Mt)^{-\frac{(3p-1)}{2}}J_{\frac{(3p-1)}{2}}(Mt)\;,
\end{equation}
which making use of the large argument expansion ($Mt\gg1$) of fractional Bessel functions~\cite{AbramowitzStegun}, can be approximated by a cosinusoidal function
\begin{equation}
\chi(t) \approx A\sqrt{\frac{2}{\pi}}(Mt)^{-\frac{3p}{2}}\cos\left(Mt-({3p/2})(\pi/2)\right)\;.
\end{equation}
The normalization constant $A$ can be fixed if we consider that the oscillatory behaviour starts just at the end of inflation, i.e. $\chi(t=0) = \chi_{\rm end} = \alpha^{-1}M_p\log(1+2/\sqrt{3})$~(\ref{chiend}). In this case,
\begin{equation}
 A = \chi_{\rm end}\,2^{\frac{1}{2}(3p-1)}\left(\frac{1}{2}(3p-1)\right)!\;,
\end{equation}
where we have made use of the limit of the Bessel functions when $Mt\ll 1$ 
\begin{equation}
J_{\frac{1}{2}(3p-1)}(Mt) \approx \frac{(Mt)^{\frac{1}{2}(3p-1)}}{2^{\frac{1}{2}(3p-1)}\left(\frac{1}{2}(3p-1)\right)!}\;.
\end{equation}
The energy and pressure densities associated to the general solution (\ref{physsolution}) are given, after averaging over several oscillations, by
\begin{eqnarray}\label{energyInflaton}
\rho_\chi \approx \left\langle\frac{1}{2}\dot\chi^2+\frac{1}{2}M^2\chi^2 \right\rangle &\approx& \frac{1}{2}M^2X^2 [\left\langle\cos^2(Mt-3\pi p/4)\right\rangle + \left\langle\sin^2(Mt-3\pi p/4)\right\rangle] = \frac{1}{2}M^2X^2\;, \\
p_\chi \approx\left\langle\frac{1}{2}\dot\chi^2 - \frac{1}{2}M^2\chi^2\right\rangle &\approx& \frac{1}{2}M^2X^2[\left\langle\cos^2(Mt-3\pi p/4)\right\rangle -\left\langle\sin^2(Mt-3\pi p/4)\right\rangle] =  0\;, 
\end{eqnarray}
with $X(t) \propto (Mt)^{-\frac{3p}{2}}$. Since the averaged pressure is negligible $p_\chi \approx 0$, then $a(t) \propto t^{p}$ 
with $p \approx 2/3$. Using this fact, the physical solution is finally expressed as
\begin{eqnarray}
\chi(t) = \frac{\chi_{\rm end}}{Mt}\sin(Mt)\;.
\end{eqnarray} 
Rewriting the previous equation in terms of the number of times the inflaton crosses zero, $j = (Mt)/\pi$, or equivalently in terms of the number of oscillations $N = j/2$, then
\begin{eqnarray}\label{chit}
\chi(t) \approx \frac{\chi_{\rm end}}{2\pi N}\sin(2 \pi N)=\frac{\chi_{\rm end}}{j\pi}\sin(\pi j)\equiv X(j) \sin(\pi j)\;,
\end{eqnarray}
Therefore, the Higgs condensate oscillates with a decreasing amplitude $X(j) \propto 1/j$. We can obtain an upper bound, $\alpha\kappa\vert\chi\vert < 0.122/N$, on the amplitude of the Higgs field after $N$ oscillations, which in terms of the correction of $\delta M^2$ to 1, see Eq.~(\ref{eqNonLinearHiggs}), implies $|\delta M^2| < 0.122, 0.0615$ or $0.0244$, after the first $N = 1, 2$ and $5$ oscillations, respectively. Thus, from the very beginning, the effective potential of the Higgs field tends very rapidly to that of a harmonic oscillator, which justifies \textit{a posteriori} the approximation $\vert \delta M^2\vert \ll 1$ used in the derivation of Eqs.~(\ref{solutionBessel}) and (\ref{chit}).

Note that if we neglect the presence of other fields and consider the Higgs-condensate as a free field only damped by the expansion rate, then we can easily estimate the number of semi-oscillations before the amplitude of the field becomes smaller than the transition value $\chi_t \sim M_p/\xi$, defined in section~\ref{Higgssector}. For $|\chi| < \chi_t$, the Higgs potential will not be anymore approximated by Eq.~(\ref{potentialEapproxII}), but rather by a quartic form $(\lambda/4)\chi^4$. This will happen when $X(j_t)/\chi_t \sim \frac{ \xi\kappa\chi_{\rm end}}{\pi j_t} < 1$, which implies 
\begin{eqnarray}
j \geq j_t \equiv \frac{\xi\log(1+2/\sqrt{3})}{\alpha\pi} \sim \mathcal{O}(10^4).
\end{eqnarray}
Therefore, if before such a moment, the Higgs field has not yet transferred efficiently its energy into other fields, then after $j_t$ semi-oscillations, the transition in the behaviour of the potential will also imply a change in the expansion rate, from a matter-like, characteristic of quadratic potentials, to radiation-like, characteristic of quartic potentials.

\subsection{Perturbative Decay of the Higgs Field}\label{perturbativereheating}

A natural reheating mechanism, given the strength of the interactions of the Higgs\index{Higgs!field}   boson with the Standard Model particles, would be a perturbative decay process of the inflaton quanta into the Standard Model particles right after the end of slow-roll. As we saw, soon after the end of inflation the effective potential for the Higgs field can be approximated by a simple quadratic potential~(\ref{potentialEapprox}). In this approximation, the masses of fermions and gauge bosons, see Eqs.~(\ref{redefmasWZ}) and (\ref{redefmassfermion}), are simply given by
\begin{equation}\label{massesapprox}
m_{f}  \simeq y_{f}\left(\frac{\alpha |\chi|}{2\xi M_P} \right)^{1/2} M_P\;, \hspace{5mm}m_{W} \simeq g_{2} \left(\frac{\alpha|\chi|}{4\xi M_P}\right)^{1/2} M_P\;,  \hspace{5mm}m_{Z} \simeq \frac{g_{2}}{\cos\theta_W} \left(\frac{\alpha|\chi|}{4\xi M_P}\right)^{1/2} M_P\;.
\end{equation}

\noindent In order to have a perturbative decay two conditions must be fulfilled :
\\

1) There should be enough phase-space in the final states for the Higgs field to decay, i.e. $M>2m_{f},m_{A}$, which will only happen when the amplitude of the Higgs field becomes smaller than a certain critial value $\chi_c$. In particular, for a decay into gauge bosons and/or fermions, in the light of Eq.~(\ref{massesapprox}), one needs
\begin{eqnarray}\label{chicritical}
\chi \gtrsim \chi_c \equiv \frac{1}{g^2}\sqrt{\frac{\lambda}{2}}M\;,
\end{eqnarray}
where $g = g_2, g_2/\cos\theta_w$ for the $W$ and $Z$ bosons, respectively, and $g = y_f$ for the fermions. When compared to the initial amplitude~(\ref{chiend}) of the Higgs field at the end of inflation, $\chi_c/\chi_{\rm end} \simeq 1/(3g^2\xi)$, we see that this critical value is much smaller than $\chi_{\rm end}$ for the gauge bosons and the top quark; of the same order for the bottom and charm quarks, and even greater for the rest of the quarks and the SM leptons.
\\

2) The Higgs decay rate $\Gamma\sim \frac{g^2}{8\pi}M$ has to be greater than the rate of expansion $H^2=\frac{\rho_\chi}{3M_P^2} \approx \frac{1}{6}(\frac{M}{M_P})^2\left(\frac{\chi_{\rm end}}{\pi j}\right)^2$,  where we have used Eq.~(\ref{energyInflaton}). Such a condition, $\Gamma > H$, can be translated into the following inequality
\begin{equation}\label{ncritic}
j \geq j_c \equiv \frac{4(\alpha\kappa\chi_{\rm end})}{g^2}\,,
\end{equation}
which defines the critical number of semi-oscillations required for this second condition to be true (again, $g = g_2, \ g_2/\cos\theta_w$ and $y_f$, for $W$, $Z$ bosons and fermions, respectively).
\\

The critical amplitude (\ref{chicritical}) below which the Higgs is allowed to decay into gauge bosons is of order $\chi_c \sim 0.1M$. As mentioned before, this amplitude is much smaller than that of the Higgs at the end of inflation, $\chi_{\rm end} \approx 0.94M_p \approx 8\times10^{5}M$. Therefore, the Higgs condensate would need to oscillate $\sim 10^{6}$ times before being able to decay through this channel. The same applies to the top quark. In the case of other fermions, due to the wide range of the Yukawa couplings, several situations can take place. For instance, the decay channel into bottom and charm quarks is opened only after a few oscillations of the Higgs, while for the rest of quarks and leptons, the decay-channel has sufficient phase space from the very end of inflation. In general, the smaller the Yukawa coupling of a given fermion species to the Higgs, the less oscillations the Higgs will go through before there is enough phase-space for it to decay into such fermion species. Notice however that  the smallness of the Yukawa coupling implies also a smaller decay rate. Consider for instance the decay of the Higgs into electrons, whose Yukawa coupling is of order $y_e \approx 10^{-6}$. From the very end of inflation, see Eq.~(\ref{chicritical}), there is phase-space in this channel for the Higgs to decay into. However, it is precisely the smallness of the electron's Yukawa coupling that allows the decay to be possible, which prevents the condition 2) to be fulfilled. The decay width is much smaller than the Hubble rate for a huge number of oscillations. Looking at Eq.~(\ref{ncritic}), we realize that the Higgs condensate should oscillate $j \sim 10^{12}$ times before the decay rate into electrons overtakes the Hubble rate.

One can check that the previous conclusions also hold for the rest of fermions of the Standard Model. When there is phase-space for the Higgs to decay into a given species, the decay rate does not catch up with the expansion rate and, viceversa, if the decay rate of a given species overtakes the expansion rate, there is no phase-space for the decay to happen\footnote{Note that the condition (\ref{chicritical}) (which prevents Higgs decay into gauge bosons and top quarks) assumes an average amplitude over a single Higgs oscillation, while smaller values are attained around the minimum of the potential when $X(t) < \chi_c$. However, when this happens the Higgs field is well inside the non-adiabatic range $|\chi| < \chi_a$~(\ref{chia2}), in which the very concept of particle (gauge bosons and top quark) is not properly defined, see section~\ref{parametricreheating}.}. 
Therefore, during a large number of oscillations, the Higgs field is not allowed to decay perturbatively in any of the Standard Model fields. Moreover, before any of those decay channels is opened, many other interesting (non-perturbative effects) will take place, as we will describe in detail in the next sections.

\subsection{Tachyonic preheating and Non-adiabatic particle production at the inflection point}\label{tachyonicreheating}

As we pointed out in section \ref{Higgssector}, the effective square mass of the Higgs field $\chi$ is negative just after the end of inflation and will be so till the inflection point. When this happens spinoidal instability takes place~\cite{Tachyonic2001,Felder:2001kt,GarciaBellido:2002} and long wavelengths quantum fluctuations $\chi_k$, with momenta $k<m_\chi$, grow exponentially. The width of the tachyonic band will be limited in our case by the point of maximum particle production, the end of inflation. At this point the effective mass $m_\chi$ takes a value 
\begin{equation}\label{massend}
m^2_{\chi}(\chi_{\rm end})= \frac{\partial^2 V(\chi)}{\partial \chi^2}\Bigg|_{\chi_{\rm end}}\approx -\frac{M^2}{30}\;,
\end{equation}
which corresponds to a maximum momentum for the tachyonic band $k_{max}=0.2 M$. This comes from
vacuum quantum fluctuations, $\chi_k(t) \propto \exp(it\sqrt{k^2+M^2_\chi}) = \exp(Mt\sqrt{1/30-(k/M)^2})$,
which grow exponentially. 

However, since the inflaton is fast rolling down the potential towards the positive curvature region, the duration of the tachyonic preheating stage is so short that the occupation 
numbers of those modes in the band do not grow significantly and the effect can be neglected. In particular, the time interval from the end of inflation till the inflection point is just $M\Delta t \approx 0.5$ and therefore, even for the fastest growing mode, $k = 0$, its growth is only $\sim e^{0.5/\sqrt{30}} \approx 1.09$. This is a negligible effect and thus, one can still consider an initial spectrum of quantum vacuum fluctuations even at the inflection point. For simplicity, all our analitycal estimations have been done ignoring this period of tachyonic instability, taking as initial conditions at the end of inflation, the amplitude of the Higgs condensate $\chi_{\rm end} \approx \frac{M_P}{\alpha}\log(1+2/\sqrt{3})$~(\ref{chiend}) and quantum vacuum fluctuations.
\\

Another physical effect before the Higgs condensate reaches the bottom of the potential for the first time, will be the particle production in the inflection point, due to the violation of the adiabaticity condition,
\begin{equation}\label{adiabinflection}
 \vert \dot\omega_k\vert \gg \vert\omega_k^2\vert \;,
\end{equation}
where the frequency of oscillation of the fluctuations is $\omega_k^2(t)=k^2+V''(\chi)$. Differenciating this and rewriting the adiabaticity condition as $\dot\omega_k\omega_k\gg\omega_k^3$, we find that only those modes within the band 
\begin{equation}\label{kinf}
k^3\ll \left|\frac{V^{'''}(\chi)\dot\chi}{2}\right|\;,
\end{equation}
are amplified. At the end of inflation $\dot H=-H^2$ which implies $\dot \chi\approx - V^{1/2}(\chi)$. Extrapolating the previous formula to the inflection point (ip) we get
\begin{equation} \label{chidot}
\dot \chi_{\rm ip} \approx \sqrt{\frac{V_0}{4}}=-\frac{\sqrt{3}M}{4\kappa}\;,
\end{equation}
which indeed seems to be  a very good approximation if we compare it with the result of a numerical solution beyond slow-roll. Inserting~(\ref{chidot}) into Eq.~(\ref{kinf}) we get
\begin{equation}\label{kinflection3}
k_{\rm ip}^3\ll \left|\frac{V^{'''}(\chi)\dot\chi_{\rm ip}}{2}\right|=\frac{H^3_{60}}{2}\;,
\end{equation}
which corresponds to a maximum excited wave number at the inflection point given by $k_{\rm ip} <  0.4 M$. Again, here the time of production is so brief that the occupation number of modes within the band is not significantly enhanced. We will have to wait until the next stage of consecutive oscillations of the Higgs field around zero for a significant production of particles.

\subsection{Instant Preheating}\label{parametricreheating}

During each oscillation of the Higgs field $\chi$, the rest of the quantum fields that couple to it will oscillate many times. Consider for instance the interaction of the Higgs field \index{Higgs!field} with the $Z$ bosons around the minimum of the potential. In this region the associated action (\ref{SSBE}) can be approximated by a trilinear interaction where the masses of the $W$ and $Z$ bosons (\ref{redefmasWZ}) are given by
\begin{equation}
\tilde{m}^2_W\simeq \frac{\alpha g^2_2 M_P}{4\xi} \vert\chi\vert\;,
\hspace{15mm}
\tilde{m}^2_Z\simeq \frac{\alpha g^2_2 M_P}{4\xi \cos^2\theta_W} \vert\chi\vert\;,
\end{equation}
which are much greater than the inflaton mass $M$ for the main part of the oscillation of $\chi$. As a result, the typical frequency of oscillation of the gauge boson is much higher than the one of the Higgs\index{Higgs!field}   field $\chi$. This implies that during most of the time the effective masses of the intermediate boson are changing adiabatically and an adiabatic invariant can be defined: the number of particles. However, for values of $\chi$ very close to zero, the adiabaticity conditions
\begin{equation}\label{adiabaticity2}
 \Big\vert \frac{ \dot{\tilde{m}}_W}{\tilde{m}^2_W}\Big\vert \ll 1\;, \hspace{15mm}\Big\vert \frac{ \dot{\tilde{m}}_Z}{\tilde{m}^2_Z}\Big\vert \ll 1\;,
\end{equation}
are violated. In such a case, there will be an inequivalence between the vacua before and after the passage of $\chi$ through the minimum of the potential, which can be interpreted as particle production~\cite{KLS94,KLS97}. In terms of the field $\chi$, the violation of~(\ref{adiabaticity2}) corresponds to the region $-\chi_a\lesssim\chi\lesssim\chi_a\;$,
\begin{equation}\label{chia}
\chi_a=\left(\frac{\xi \vert\dot\chi(t)\vert^2}{\alpha g^2 M_P}\right)^{1/3} \;, 
\end{equation}
where, from now on, $g = g_2,\,\, g_2/\cos\theta_W$ for the $W$ or $Z$ bosons. Only outside this region, the notion of particle makes sense and an adiabatic invariants can be defined. Taking into account that and approximating the velocity of the field around zero as $\dot \chi(j) \approx M\,{\chi_{\rm end}\over \pi j}=MX(j)$, see Eq. (\ref{chit}), 
the general expressions (\ref{chia}) can be approximated as 
\begin{equation}\label{chia2}
\chi_a(j)=\left(\frac{\xi M^2\vert\chi_{\rm end}\vert^2}{\alpha j^2 g^2 \pi^2 M_P}\right)^{1\over3} = \left(\frac{\lambda\pi}{3\xi\log(1+2/\sqrt{3})}\right)^{1\over3}\frac{j^{1\over3}}{g^{2\over3}}\,X(j)\,.
\end{equation}
Note that the previous regions are indeed very narrow compared to the amplitude of the oscillating Higgs, $\chi_a \sim 10^{-2}j^{1/3}X(j)
$. 
Therefore, the particle production that takes place in that region happens within a very short period of time as compared to the inflatons' oscillation period $T=2\pi/M$,
\begin{equation}\label{timea}
\Delta t_a(j)\sim \frac{2\chi_a}{\vert\dot\chi\vert}\sim 10^{-2}\,j^{1/3} \; M^{-1} \ll T\;, 
\end{equation}
Notice indeed that different values of $\lambda$ do not change appreciably the above conclusions about the smallness of the non-adiabatic regions. Given the weak dependence of $\Delta t \propto j^{1/3}$, many semi-oscillations ($\sim 10^3$) will pass before the fraction of time spent in the non-adiabatic zone will increase from a 1\% to a 10\%, as compared with the period of oscillations. This holds also independently of the species, $W$ or $Z$ bosons.

Moreover, despite the smallness of $\chi_a$ as compared to the amplitude $X(j)$, it is important to note that the field range corresponding to the region of non-adiabaticity still is much greater than those critical regions defined in section~\ref{Higgssector}. In particular, let us recall that there is a field value, $\chi_t \sim M_P/\xi$, below which there is a transition of the effective potential from a quadratic to quartic behaviour. However, this is well inside the region of non-adiabaticity, $\chi_t \ll \chi_a$, as we emphasized before. Moreover, there is also an interval of Higgs field values, $|\chi| < \chi_c$ for which the Higgs perturbative decay into $W$, $Z$ and top quarks can occur [see Eq.~(\ref{chicritical}) in section~\ref{perturbativereheating}], which nevertheless is also much smaller than the non-adiabaticity interval, $\chi_c \ll \chi_a$. 

We will now discuss the non-perturbative creation of particles in the non-adiabatic region. This production is formally equivalent to the quantum mechanical problem of a particle scattering in a periodic potential. In the case under consideration the equations of motion for the fluctuations of each gauge field with a given polarization will be given by $W_k''+(k^2/a^2+\tilde m_W^2)W_k = 0$, and the corresponding one for the $Z$-fluctuations. Expanding Eq.~(\ref{chit}) around the $j$-th zero at time $t_j = \pi j$, the evolution equation of the fluctuations can be approximated as 
\begin{eqnarray}\label{fluctuations}
&& W_k'' + \left(\frac{k^2}{a^2}+\frac{\alpha g_2^2M_p\chi_{\rm end}|\sin(M(t-t_j))|}{4\pi j\,\xi}\right)W_k = 0\;,\\
\label{fluctuations2}
&& Z_k'' + \left(\frac{k^2}{a^2}+\frac{\alpha g_2^2M_p\chi_{\rm end}|\sin(M(t-t_j))|}{4\pi j\,\xi \cos_W^2}\right)Z_k = 0\;.
\end{eqnarray}
Notice that around the zeros of the inflaton the sinusoidal behaviour $|\sin(M(t-t_j))|$ can be very well approximated by  $|\sin(M(t-t_j))| \approx  M|t-t_j| \equiv \tau$, which allows us to rewrite Eqs.~(\ref{fluctuations}) and (\ref{fluctuations2}) as a Schr\"odinger-like equation like
\begin{eqnarray}\label{schrodinger}
-W_k'' - \frac{q_W}{j}|\tau|W_k = K^2 W_k\;, \hspace{15mm} -Z_k'' - \frac{q_Z}{j}|\tau|Z_k = K^2 Z_k\;, 
\end{eqnarray}
where primes denote derivatives with respect to the rescaled time $\tau = Mt$, $K$ is the rescaled momentum $K \equiv \frac{k}{aM}$ and
\begin{eqnarray}\label{qfactor}
&& q_W = \cos\theta_W^2 q_Z= \frac{g_2^2\alpha\kappa\chi_{\rm end} }{4\pi\xi}\left(\frac{M_p}{M}\right)^2 = \frac{3g_2^2\xi\,\alpha\kappa\chi_{\rm end}}{4\pi\lambda}\;, 
\end{eqnarray}
are the usual resonance parameters~\cite{KLS97}. Each time the inflaton crosses zero can be interpreted therefore as the quantum mechanical scattering problem of a particle crossing an inverted triangular potential. Let $T$  and $R = 1- T$, for either $W$ or $Z$, be the transmission and reflection probabilities for a single scattering in this periodic triangular barrier. The number of particles just after the $j$-th scattering, $n_{k}(j^+)$, in terms of the previous number of particles $n_{k}(j^-)$ just before that scattering, can be written as~\cite{KLS94,KLS97}
\begin{eqnarray}\label{occupnum}
n_{k}(j^+) = (T^{-1}_{k}(j)-1) + (2T^{-1}_{k}(j)-1)n_{k}(j^-) +2\cos\theta_j {\sqrt{T^{-1}_{k}(j)\left(T^{-1}_{k}(j)-1\right)}}\sqrt{n_{k}(j^-)\left(n_{k}(j^-)+1\right)}\;,
\end{eqnarray}
where $\theta_j$ are some accummulated phases at each scattering, that we will discuss later on in section~\ref{stimulateddecays}, since they will not play any role in the following discussion of this section. The inverse of the transmission probability for the $j$-th scattering can be expressed as \cite{GalindoPascualbook}
\begin{eqnarray}\label{Transmision}
T^{-1}_{k}(j) = 1 + \pi^2\left[{\rm{Ai}}\left(-x_j^2\right){\rm{Ai}}'\left(-x_j^2\right) + {\rm{Bi}}\left(-x_j^2\right){\rm{Bi}}'\left(-x_j^2\right)\right]^2\,,
\end{eqnarray}
with $x_{j} \equiv K/(q/j)^{1/3}$ and $\rm{Ai}(z),\rm{Bi}(z)$ the Airy functions. Note that we have used the Wronskian normalization, $\rm{Ai}(z)\rm{Bi}'(z) - \rm{Bi}(z)\rm{Ai}'(z) = \pi^{-1}$. 

\begin{figure}
\centering
\includegraphics[width=8cm,height=6cm]{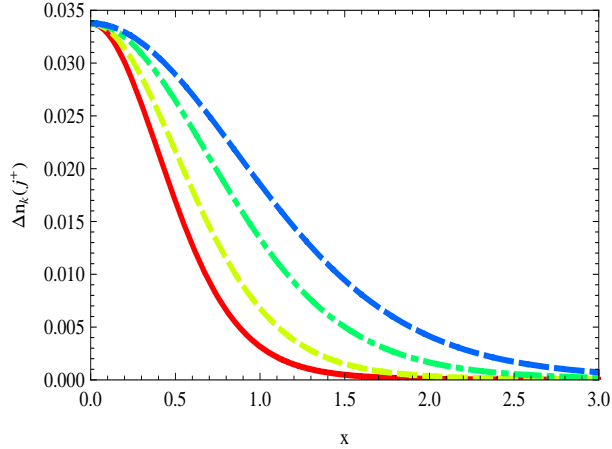}
\caption{Spectral distributions~(\ref{deltaN}) for the gauge bosons created in a single zero crossing through the first term of Eq.~(\ref{occupnum}), calculated after $j = 1, 2, 5$ and $10$ oscilations (from left to right). The horizontal axis represents $x \equiv k/Mq^{1/3}$, so $x = 1$ is the typical width of the band of momenta of particles created at the first scattering. For later times, the distributions broaden out to greater momenta, since the argument of Eq.~(\ref{deltaN}), $x_j$ behaves as $\propto j^{-1/3}$. The typical momenta of the distribution agree with the one calculated in section \ref{parametricreheating}.}
 \label{triangularT}
\end{figure}

Consider the situation, $n_{k}(j^-) \ll 1$, which is certainly true in the first scattering $j = 1$, or can happen for $j > 1$ if the previously produced gauge bosons have fully decayed into fermions. In such a case,
\begin{equation}\label{deltaN}
\Delta n_{k}(j^+) \approx T^{-1}_{k}(j)-1\;,
\end{equation}
where we have retained only the first term of Eq.~(\ref{occupnum}). This corresponds to the spontaneous particle creation of $W$ and $Z$ bosons each time the Higgs crosses zero and, therefore, tells us about the number of particles of these species that are created in each zero-crossing. The momenta distribution is shown in Fig.~\ref{triangularT}. In particular, the total number of produced particles of a given species with a given polarization, just after exiting the non-adiabatic region around the $j$-th zero-crossing, can be obtained as
\begin{equation}
\Delta n(j^+) =\frac{1}{2\pi^2\,a_j^3}\int_0^\infty dk\,k^2\,\Big[T^{-1}_{k}\left(j\right)-1\Big] =\frac{q}{2j}\,\mathcal{I}\,M^3\,,
\end{equation}
with $\mathcal{I} = \int_0^\infty\left[{\rm Ai}(-x^2){\rm Ai}'(-x^2) + {\rm Bi}(-x^2){\rm Bi}'(-x^2)\right]^2x^2dx \approx 0.0046$ and $q$ the resonant parameters given by Eq.~(\ref{qfactor}). Thus, the only difference between the number of $W$ and $Z$ bosons produced is simply encoded in the different resonance parameter, $q_W \propto g_2^2$ and $q_Z \propto g_2^2/\cos^2\theta_W$, respectively.  Notice that the effect of the non-perturbative production of these particles is proportional to the coupling square. If the couplings of the Higgs field to the gauge bosons were not so large ($g_2 \sim 0.5, \cos^{-1}\theta_W \sim 1.4$), then their produccion would be very supressed. Strictly speaking the previous analysis is just valid for gauge bosons. The production of fermions through this mechanism is different and more involved than for bosons. Nevertheless, if the effect is, as expected, proportional to the Yukawa coupling squared \cite{GarciaBellido:2001cb} then 
only the top quark production would be non-negligible. 

After the passage through the minimum of the effective potential the number of $W$ and $Z$ particles remains (almost) constant while their masses grow when the field $\chi$ increases. The $W$ and $Z$ bosons tend to decay into fermions in a time $\Delta t\sim \langle\Gamma^E_{W,Z}\rangle_j^{-1}$, where $\Gamma^E_{W,Z}$ are given by Eq.~(\ref{widthWE}), 
while $\langle \cdot \rangle_j$ represents a time average between the $j$- and the $(j+1)$-th scatterings. Given the time-dependence 
of the $\chi$ field~(\ref{chit}), the typical time of decay turns out to be $\Delta t \simeq 0.64\, j^{1/2}\, M^{-1}$ for the $Z$ bosons and a bit bigger, $\Delta t \simeq 1.55\, j^{1/2}\, M^{-1}$, for the $W$ bosons, as was expected. This implies that in a semiperiod, $T/2 = \pi M^{-1}$, the non-perturbatively produced gauge bosons at the $j$-th scattering, decay significantly before the next scattering takes place, at least for the first scatterings. However, as the amplitude of the Higgs field decreases with time due to the expansion of the Universe, the probability of decay of the gauge bosons, see Eq.~(\ref{widthWE}), 
becomes smaller and smaller as time goes by. This explains the $j^{1/2}$ behaviour, which essentially means that after a certain number of oscillations, the number of produced fermions through the perturbative decays per semi-oscillation, will become eventually negligible.

In the Jordan frame the Standard Model presents its usual form and the fermions produced in the decay of the $Z$ and $W$ bosons are mainly relativistic. Since both momenta and masses transform in the same way under a change of conformal frame, if a gauge boson is allowed to decay into a pair of fermions in the Jordan frame, it will also be able to decay in the Einstein frame. Therefore, the relation between the typical momenta and masses of those fermions (F) and gauge bosons (W,Z) in the conformally transformed frame, is simply given by
\begin{equation}\label{deltaWZ}
2(\tilde{k}^2_F + \tilde{m}^2_F) = \tilde{k}_W^2 + \tilde{m}_W^2 \;,\hspace{1cm}
2(\tilde{k}^2_F + \tilde{m}^2_F) = \tilde{k}_Z^2 + \tilde{m}_Z^2 \;.
\end{equation}
In terms of the field $\chi$, the previous equations can be rewritten as
\begin{eqnarray}
\label{deltaWE}
\frac{\tilde{k}^2_F}{\tilde{m}^2_F} &=& \frac{1}{y_F^2}\left(\frac{\xi \tilde{k}_W^2}{M_P^2(e^{\alpha\kappa\vert\chi\vert}-1)}+\frac{g_2^2}{4}\right)-1\;,\\
\label{deltaZE}
\frac{\tilde{k}^2_F}{\tilde{m}^2_F} &=& \frac{1}{y_F^2}\left(\frac{\xi \tilde{k}_Z^2}{M_P^2(e^{\alpha\kappa\vert\chi\vert}-1)}+\frac{g_2^2}{4\cos^2\theta_W}\right)-1\;, \end{eqnarray} 
for the $W$ and $Z$ fields respectively. Note that the relativistic or non-relativistic nature of a given particle is something intrinsic to the particle and should not depend on the conformal frame. 
As expected the transitions $Z\rightarrow \bar{t}t$, $W\rightarrow t b$ are not allowed in the Einstein frame. For the rest of quarks $\tilde k_F^2 \gg \tilde m_F^2$, which implies that all the fermions produced in the decay of the $W$ and $Z$ bosons are clearly relativistic, as happened in the Jordan frame.
The total number density of gauge bosons $n(j^+)$ present just after the $j$-th crossing will decay exponentially fast  until the next crossing, due to the perturbative decay into fermions. Therefore the total number density just previous to the $(j+1)$-th zero crossing, $n((j+1)^-)$, is given by
\begin{eqnarray}
n((j+1)^-) = n(j^+)e^{-\int_{t_j}^{t_{j+1}} \Gamma dt} = n(j^+)e^{-\langle\Gamma\rangle_j\frac{T}{2}}\;.
\end{eqnarray} 
The number of fermions produced between those two scatterings, $\Delta n_{F}(j)$, is simply given by
\begin{eqnarray}
\Delta n_{F}(j) = 2\times3\times\left[n_Z(j^+)(1-e^{-\langle\Gamma_Z\rangle_j\frac{T}{2}}) + 2\,n_W(j^+)(1-e^{-\langle\Gamma_W\rangle_j\frac{T}{2}})\right]\;,
\end{eqnarray}
where the factor $2\times3$ takes into account that each gauge boson can have one out of three polarizations and decay into two fermions, while the extra factor $2$ in front $n_W$, accounts both for the $W^+$ and $W^-$ decays. The averaged value of the decay widths in the previous expressions can be estimated, see Eqs.~(\ref{widthWE}), as 
\begin{eqnarray}
\left\langle\Gamma_{Z\rightarrow all}\right\rangle_j &=& \left(\frac{g_2 }{\cos\theta_W}\right)^3\frac{M_P\,{\rm Lips}}{16\pi\sqrt\xi}\left\langle (1-e^{-\alpha\kappa|\chi|})^{1/2} \right\rangle_{{j}}  \equiv \frac{2\gamma_Z}{T} F(j)\,,\\ 
\left\langle\Gamma_{W\rightarrow all}\right\rangle_j &=& \frac{3\cos^3\theta_W}{2{\rm Lips}}\left\langle\Gamma_{Z\rightarrow all}\right\rangle_j\equiv \frac{2\gamma_W}{T} F(j)\;,
\end{eqnarray}
where $T=2\pi/M$ is the typical oscillation period and we have defined
\begin{eqnarray}\label{Fn}
F(j) \equiv  \left\langle \Big(1-e^{-\alpha\kappa|\chi|}\Big)^{1/2} \right\rangle_{j} 
= \left(\frac{1}{\pi}\int_{j\pi}^{(j+1)\pi}{dx\left[1-\left(1+\dfrac{2}{\sqrt{3}}\right)^{-\left|\frac{\sin x}{x}\right|}\right]^{\frac{1}{2}}}\right) \approx 0.3423\frac{1}{\sqrt{j}}\,.
\end{eqnarray}
Note that the last approximated equality is simply a (good) fit to $F(j)$ for all $j$. The constants $\gamma_Z,\gamma_W$ are just numerical factors depending of the parameters of the model and the decaying species,  
\begin{eqnarray}\label{smallGamma}
\gamma_Z = \left(\frac{g_2}{\cos\theta_W}\right)^{3}\frac{\sqrt{3}\xi^{1/2}}{16\lambda^{1/2}}\,{\rm Lips} \approx 14.23\lambda^{-\frac{1}{4}}\,, \hI \gamma_W \equiv \frac{3\cos^3\theta_W}{2{\rm Lips}}\gamma_Z \approx  5.91\lambda^{-\frac{1}{4}},
\end{eqnarray}

Using the notation, $E_{F_{\rm Z}}{(j)}$ and $E_{F_{\rm W}}{(j)}$ for the mean energy of the fermions produced between $t_j$ and $t_{j+1}$, from the decay of $Z$ or $W$ bosons, respectively, then we find
\begin{eqnarray}\label{fermionsEnergy}
&&E_{F_{\rm Z}}{(j)} \equiv \left\langle \sqrt{k_F^2 + m_F^2} \right\rangle_j \approx \langle k_F \rangle_j \approx \frac{1}{2}\langle m_Z \rangle_j \approx \frac{g_2}{4\xi^{1/2}\cos\theta_W}\,F(j)\,M_p\,,\\
&&E_{F_{\rm W}}{(j)} \equiv \left\langle \sqrt{k_F^2 + m_F^2} \right\rangle_j \approx \langle k_F \rangle_j \approx \frac{1}{2}\langle m_W \rangle_j \approx \frac{g_2}{4\xi^{1/2}}\,F(j)\,M_p\,,
\end{eqnarray}
where we have used the fact that the produced fermions are relativistic, see Eqs~(\ref{deltaWE}) and (\ref{deltaZE}), while the gauge bosons are non-relativistic, see Eq.~(\ref{gaugeBosonsMomenta}) in section~\ref{stimulateddecays}.
\\

Let us work now under the following hypothesis: we will consider that the perturbative decay of the gauge bosons into fermions is sufficiently effective, such that the gauge bosons do not accumulate significantly. This amounts to neglect initially a potential effect of parametric resonance. This is of course a rough approximation, which is only valid for the first oscillations, where $e^{-\gamma\,F(j)} \ll 1$. Numerically, after $j = 1, 2, 10, 15$ and $20$ zero-crossings, the 99.5\%, 98.5\%, 94.2\%, 87.4\%, 81.9\%, 77.4\% respectively, of the produced $Z$ particles have decayed into fermions (and a similar though smaller fraction of the $W$ bosons). This implies that there will be always a remnant of the gauge bosons produced at each scattering, that will not decay in one semi-period of the inflaton's oscillation. Let us neglect this for the time being, therefore ignoring the possibility of having parametric resonance, and estimate the energy transferred simply through the perturbative decay into fermions, during the first oscillations.

In particular, the energy density of those fermions produced after the first scattering, averaged over the first semi-oscilation between $Mt = \pi $ and $Mt = 2\pi$, will be
\begin{eqnarray}
\Delta \rho_F{(1)} &\sim& 6\left[\Delta n_Z(1)E_{F_{\rm Z}}{(1)} + 2\Delta n_W(1)E_{F_{\rm W}}{(1)}\right] = \varepsilon\left(\frac{1}{2}M^2\chi_{\rm end}^2\frac{1}{\pi^2}\right)\, F(1)\,,
\end{eqnarray}
with
\begin{eqnarray}\label{factor}
\varepsilon \equiv \frac{3^{3/2}\pi\alpha^2(2+\cos^{-3}\theta_W)\mathcal{I}g_2^3}{8\lambda^{1/2}\xi^{1/2}(\alpha\kappa\chi_{\rm end})} \approx 3\times10^{-5}\lambda^{-3/4}\,.
\end{eqnarray}
The energy density of the inflaton, evaluated at the maximum amplitude of the first semi-oscillation, is given (\ref{energyInflaton}) by
\begin{eqnarray}
\rho_\chi{(1)} \approx \frac{1}{2}M^2\chi_{\rm end}^2\left(\frac{2}{3\pi}\right)^2\,.
\end{eqnarray}
Therefore, the ratio between the energy density of the fermions and of the inflaton, at that moment, $Mt \approx 1.5\pi$, is 
\begin{eqnarray}
\epsilon{(1)} \equiv \frac{\rho_F{(1)}}{\rho_\chi{(1)}} = \frac{\varepsilon F(1)}{(2/3)^2} \approx 2\times10^{-5}\lambda^{-3/4}\,,
\end{eqnarray}
which means that initially, for e.g.~for $\lambda = 0.4$, only a $\sim 0.004\%$ of the inflaton's energy has been transferred to the fermions. Thus, the so called Instant Preheating mechanism (\cite{InstantFKL98}) results frustrated here, because in order to make it work efficiently, the couplings of the theory must be really fine-tuned, in such a way that a significant fraction of the energy of the inflaton was transferred (in the first semi-oscillation) to the decay products of the bosons to which the inflaton is coupled. Moreover, in the instant preheating scenario, the produced fermions must be non-relativistic while the effective behaviour of the background inflaton should be effectively mimicing that of relativistic matter (like $e.g.$ in $\lambda\chi^4$ models). Only in this case it would be guaranteed that the remnant energy of the inflaton would decay faster that that of the fermions, thanks to the extra supression factor $1/a$ due to the expansion of the Universe. If the inflaton would  effectively behave as non-relativistic matter and the produced fermions were relativistic, the energy of the inflaton could again overtake very soon that of the fermions, because now the fermion's energy  would decrease faster than that of the background. That is, precisely, the situation we have in the scenario under discussion. Even if we had found that $\epsilon{(1)} \sim \mathcal{O}(1)$, the relativistic nature of the fermions and the non-relativistic effective behaviour of the Higgs oscillations, would have prevented the Universe to instantaneously reheat at that point.

One could hope that after a certain number of oscillations, let us say $j_p$, that ratio would grow up to a value $\epsilon{(j_p)}\sim \mathcal{O}(1)$. The succesively produced fermions, generated each semi-oscillation through the perturbative decay of the (non-perturbatively produced) $W$ and $Z$ bosons, could perhaps accumulate a sufficiently ammount of energy that could finally equal that of the Higgs condensate. This seems not totally unresonable because the total energy stored in the Higgs decreases with the expansion of the Universe as $\rho_\chi \propto 1/j^2$, see Eq.~(\ref{energyInflaton}), such that the total amount of energy that we would require to transfer to the fermions would be less and less. Moreover, the number of fermions would only increase as time goes on, so one keeps adding energy each semi-oscillation in the form of new produced fermions. Thus, these two effects would contribute to the increment of the ratio of the energy between the fermions and the Higgs. On the other hand, the relativistic nature of the fermions and the decrease of their production rate with the expansion, as $\Gamma \propto 1/\sqrt{j}$, would tend to decrease such a ratio. Therefore, one must put together all these competing effects in order to obtain the evolution in time of the energy transferred from the Higgs to the fermions. To do this, we will assume, both for simplicity as well as for trying to make this mechanism more efficient, that since the gauge bosons do not accumulate significantly, only the first term of Eq.~(\ref{occupnum}) should be considered, such that $W$ and $Z$ bosons are only produced through spontaneous creation at each zero-crossing. 

In this case the averaged energy density of the fermions produced between $t_j$ and $t_{j+1}$ will be given by
\begin{equation}
\Delta \rho_F{(j)} \sim 6\left[\Delta n_Z{(j)}(1-e^{-\gamma_Z F(j)})E_{F_{\rm Z}}(j) + 2\Delta n_W{(j)}(1-e^{-\gamma_WF(j)})E_{F_{\rm W}}(j)\right] = \varepsilon\left(\frac{1}{2}M^2\chi_{\rm end}^2\frac{1}{\pi^2}\right)\,\frac{F(j)}{j}\,\Upsilon(j)\,,
\end{equation}
where $\varepsilon$ is given by Eq.~(\ref{factor}) and we have defined
\begin{eqnarray}
\Upsilon(j) \equiv \left(1-\frac{(1+2\cos^3\theta_We^{-(\gamma_W-\gamma_Z)F(j)})}{(1+2\cos^3\theta_W)}\,e^{-\gamma_Z F(j)}\right)\,.
\end{eqnarray}
Then the ratio between the energy of the fermions to the Higgs condensate at the $j$-th zero crossing, finally reads 
\begin{eqnarray}\label{succesiveInstant}
\varepsilon{(j)} \equiv \frac{\rho_F{(j)}}{\rho_\chi{(j)}} \approx \varepsilon\frac{\left(j+\frac{1}{2}\right)^2}{j}\sum_{i=1}^{j} F(i)\Upsilon(i)\left(\frac{i}{j}\right)^\frac{5}{3} =  \,\varepsilon\,G(j)\frac{\left(j+\frac{1}{2}\right)^2}{j}\,,
\end{eqnarray}
where
\begin{eqnarray}
G(j) \equiv \sum_{i=1}^{j} F(i)\Upsilon(i)\left(\frac{i}{j}\right)^{5/3}\,.
\end{eqnarray}
Note that the strength of the effect, i.e. the amplitude of $\varepsilon(j)$, is modulated by the gauge couplings through $\varepsilon \propto g_2^3$ so, even in this case in which the SM gauge couplings of the Higgs to the vector bosons are quite big ($g_2^2 \sim 0.3$), that does not help to transfer sufficient energy initially. As mentioned before, if we could apply this formalism to the production of fermions at each Higgs' zero crossing, by substituying the gauge couplings with the Yukawa ones, we would obtain even a more ridiculous production of particles (except perhaps for the top quarks). Of course, the question of fermionic preheating at each zero crossing deserves more investigation and we will address it in a future publication.

The numerical values of the ratio $\varepsilon{(j)}$ after e.g. $j = 1, 2, 5, 10, 15$ and $20$ semi-oscillations, for e.g.~for $\lambda = 0.4$, are respectively $\varepsilon{(j)}[\times10^5] \sim 3.90, 5.97, 11.82, 37.26, 65.04$ and $97.59$. For different values of $\lambda$, these numbers do not change significantly. Thus, we see that the transferred energy from the Higgs field to the fermions through the gauge bosons is generically a very slowly growing function. After $20$ crossings the transferred energy is still only $\sim 0.03\%$ of the Higgs energy at that time. Therefore, we clearly see that this \textit{succesive Instant Preheating} mechanism is not efficient enough as to rapidly reheat the Universe. If we consider that the former formalism~(\ref{succesiveInstant}) is valid up to an arbitrary number of oscillations, then we can estimate the number $j_p$ of semi-oscllations required to achieve $\varepsilon(j) \sim \mathcal{O}(1)$. Equating $\varepsilon(j)$ to 1 in Eq.~(\ref{succesiveInstant}), we obtain $j_p \sim \mathcal{O}(10^4)$. However, much earlier than that, parametric resonance effects should be considered, see section~\ref{stimulateddecays}.

In other words, notice that 
we have neglected the presence of those $Z$ and $W$ bosons that did not decay into fermions in each semi-oscillation. The occupation number of the bosons produced at the bottom of the potential is not simply generated by the first term of Eq.~(\ref{occupnum}), but rather by the rest of the terms in Eq.~(\ref{occupnum}), which indeed give rise to the phenomena of resonant production of bosons. Taking this into account will have very interesting consequences. The number density of the bosonic species will grow exponentially fast and thus will also transfer energy into the fermions exponentially rapidly. We must therefore develop a mixed formalism that takes into account the two competing effects: that of parametric resonant production of bosons versus the effect of their perturbative decay into fermions. It is crucial to note that while the perturbative decay does not transfer enough energy (as we have just seen), the fact that those bosons disappear will have very important consequences for the development of the resonant effect. In particular, the resonance will not become effective from the beginning of the oscillations of the inflaton right after inflation, as usually assumed, but only after the inflaton has already performed a significant number of oscillations.

\section{Combined Preheating: mixed Parametric Resonance and Perturbative Decays}\label{stimulateddecays}

Let us now analyze how the occupation number of the $Z$ and $W$ bosons grow if we consider the effect of all the terms in Eq.~(\ref{occupnum}). The production of gauge bosons will also occur, as before, in a very short interval of time~(\ref{timea}) when the Higgs condensate crosses around zero, violating then the adiabaticity conditions ~(\ref{chia2}), $|\chi| < \chi_a$. In the large occupation limit $n_k \gg 1$, the first term in Eq.~(\ref{occupnum}) can be neglected and therefore the spectra number density of the produced gauge bosons just after the $j$-th scattering is given by
\begin{eqnarray}
n_k(j^+) \approx \left( (2T^{-1}_k(j)-1) - 2\cos\theta_j\sqrt{T^{-1}_k(j)(T^{-1}_k(j)-1)}\right)n_k(j^-)\,,
\end{eqnarray}
which indicates the spectral number density $n_k(j^+)$ just after the $j$-th scattering, in terms of the spectral number density $n_k(j^-)$ just before such scattering. Since the interval between successive scatterings is $M\Delta t = \pi$, we can define naturally a growth (Floquet) index $\mu_k(j)$ as~\cite{KLS94,KLS97}
\begin{eqnarray}\label{expGrowth}
n_k(j^+) \approx n_k(j^-)e^{2\mu_k(j)M\Delta t} = n_k{(j^-)}e^{2\pi\mu_k(j)}\,.
\end{eqnarray}
Comparing formulas, we obtain
\begin{eqnarray}\label{floquet}
\mu_k(j) \approx \frac{1}{2\pi}\log\left( (2T^{-1}_k(j)-1) - 2\cos\theta_j\sqrt{T^{-1}_k(j)(T^{-1}_k(j)-1)}\right)\,.
\end{eqnarray}
The $\theta_j$ are some accummulated phases at the $j$-th scattering, which can indeed play a very important role, since they can enhance ($\cos\theta_j < 0$) or decrease ($\cos\theta_j > 0$) the effect of production of particles at each scattering.

\begin{figure}[t]
\hspace{-0.55cm}\includegraphics[width=7cm,height=5cm]{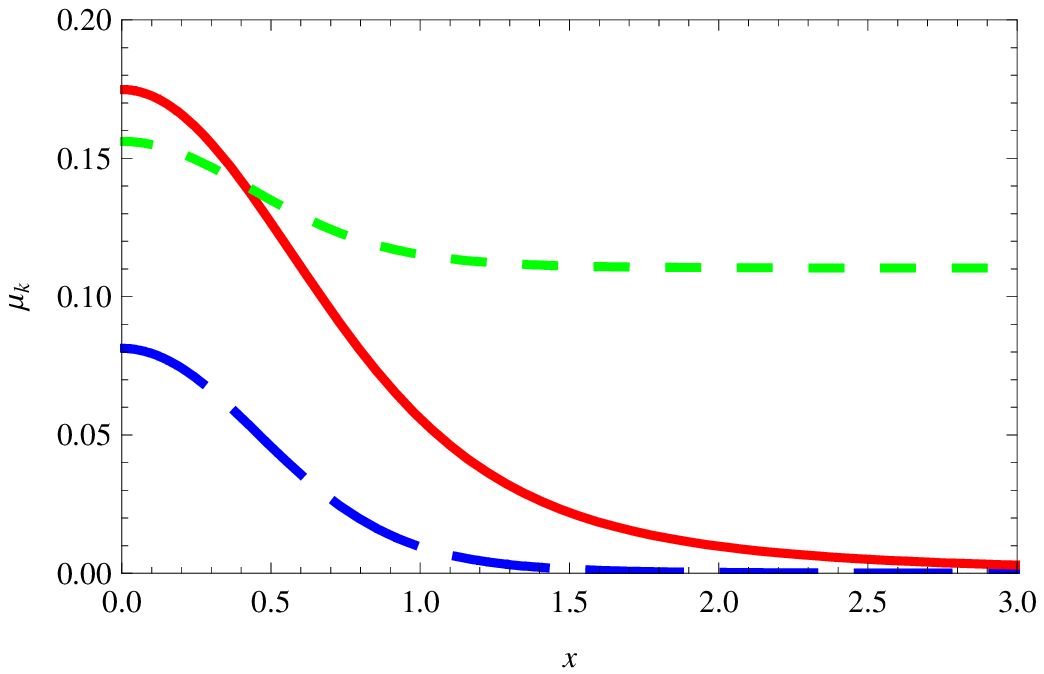}
\hspace{2cm}\includegraphics[width=7cm,height=5cm,angle = 0]{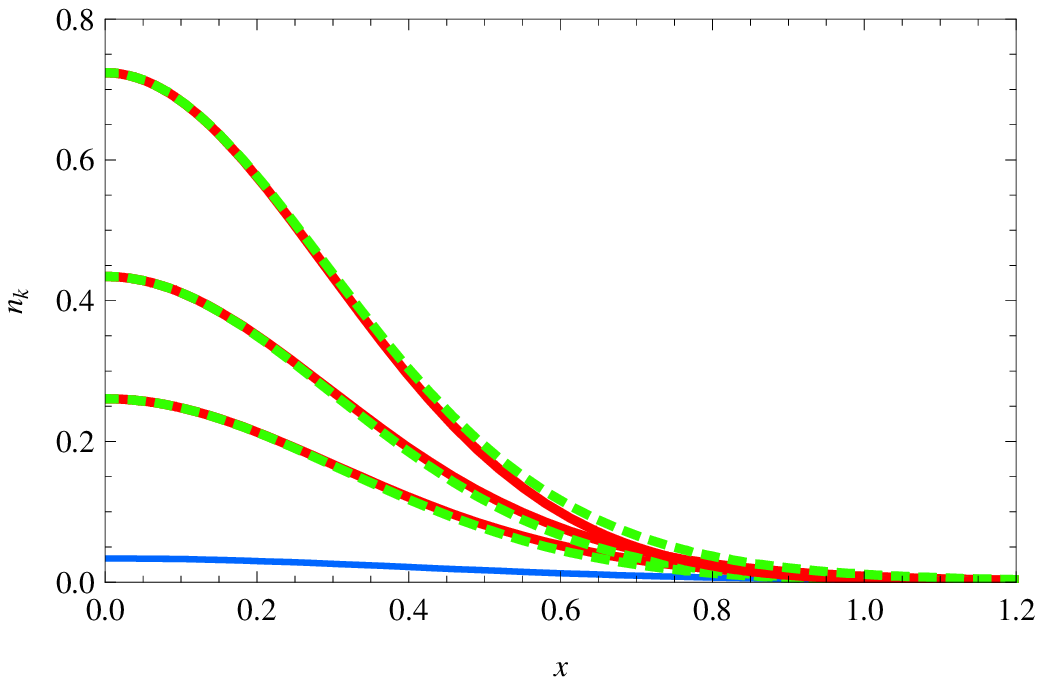}
\caption{Left: The Floquet index for a given polarization of the $W$ and $Z$ bosons as a function of the variable $x_j = k/
k_*{(j)}$. Here we show the maximum (continuous red), the average (short dashed green) and the typical (long dashed blue) indices. Right: The initial spectral distribution $n_k(1^+)$ (lower blue curve) and the Gaussian approximation $n(j^+\geq2)$~(\ref{occupCombPre}) for different $j's$ greater than 2 (rest of the curves), 
describing the resonant behaviour. The approximation is so good that it is hard to distinguish it from the real curve, presenting small deviations just on the tail. The horizontal axis is $x = k/
k_*{(1)}$ and the curves correspond to different $j$'s. It is clearly distinguishable the fact that only the range $x < 1$ ($k < 
k_*{(1)}$) is filtered and therefore excited through parametric resonance, no matter if $j \gg 2$.}
\label{mubehaviour}
\end{figure}

Depending on the phases, we can consider the following cases: The typical behaviour of the Floquet index, for $\cos\theta=0$, 
\begin{eqnarray}
\mu_k^{(\rm typ)} = {1\over2\pi}\log\left(2T^{-1}_k-1\right)\,,
\end{eqnarray} 
the maximum index, achieved for $\cos\theta=-1$, given by
\begin{eqnarray}
\mu_k^{({\rm max})} = {1\over\pi}\log\left( \sqrt{T^{-1}_k} + \sqrt{T^{-1}_k-1}\right)\,,         \end{eqnarray}
and the average index over an oscillation, obtained as 
\begin{eqnarray}
\mu_k^{(\rm av)} = {1\over2\pi}\int_0^{2\pi} \mu_k(\theta)\,d\theta = \frac{1}{2\pi}\log\left(2T^{-1}_k\right)\,.
\end{eqnarray}
All these possibilities are shown in Fig.~\ref{mubehaviour}, as a function of $x \equiv K/(q/j)^{1/3}$, where $q$ are the resonant parameters (\ref{qfactor}) for the $Z$ and $W$ bosons, while $x_j$ is the natural argument of the transmission probability scattering functions~(\ref{Transmision}).

As explained in Ref.~\cite{KLS97}, when $\Delta\theta_j \equiv \theta_{j+1} - \theta_{j} \gg \pi$, the effect of resonance will be chaotic, being then the phases essentially random at each scattering. For instance, using the effective frequencies of the fluctuations (\ref{fluctuations}) of the W field, these phases can be estimated, for the relevant range of momenta, as follows
\begin{eqnarray}
\Delta\theta_j = \int_{t_j}^{t_{j+1}}dt\sqrt{K^2+{\tilde m}^2_{W}} \approx
\frac{g_2\pi\sqrt{3\xi}}{2\sqrt{\lambda}}F(j) \sim \mathcal{O}(10^{-2})j^{-1/2}\,,
\end{eqnarray}
where $F(j)$ was defined in Eq.~(\ref{Fn}) and we have neglected $K^2$ versus $\tilde m_{W}^2$ in the second equality, since, as will be justified later~(\ref{gaugeBosonsMomenta}), the produced bosons are non-relativistic. Comparing the above formula with $\pi$, we see that the end of the stochastic behaviour will occur after $\sim (2-5)\times10^{3}$ zero crossings, depending on $\lambda$. For the case of the $Z$ boson the previous estimation of the end of the stochastic resonance is modified by a factor $(\cos\theta_W)^{-1}\approx {\cal O}(1)$, being thus the result essentially unaffected. Therefore, since for the first thousand of oscillations of the Higgs, the accummulated phases of the fluctuations of the gauge bosons will be chaotic, we will average out the phases and work with $\mu_k^{({\rm av})}$.

On the other hand, the perturbative decay of the produced vector bosons occurs precisely just between two successive Higgs zero-crossings, $n((j+1)^-) = n(j^+)\exp(-\gamma\,F(j))$,
where $F(j)$ is given by Eq.~(\ref{Fn}) and $\gamma=\gamma_Z,\gamma_W$, see Eq. (\ref{smallGamma}). Taking into account Eq.~(\ref{expGrowth}) and Eq.~(\ref{floquet}) we can express the number of gauge bosons just after the $(j+1)$-th scattering in terms of the number just after the previous one
\begin{eqnarray}
&& n_{k}((j+1)^+) = n_{k}((j+1)^-)e^{2\pi\mu_{k}(j+1)} = n_{k}(j^+)e^{-\gamma\,F(j)}e^{2\pi\mu_{k}(j+1)}\;,
\end{eqnarray} 
Applied recursively, this formula allows us to obtain the occupation number for each species and polarization, just after the $(j+1)$-th scattering in terms of the initial abundances $n_k(1^+)$,
\begin{eqnarray}\label{formulaCrecimiento}
&& n_{k}((j+1)^+) = n_{k}{(1^+)}\exp\Big[-\gamma\sum_{i=1}^jF(i)\Big]\,\exp\Big[2\pi\sum_{i=1}^{j}\mu_{k}(i+1)\Big] \;. 
\end{eqnarray}
The initial abundances are, of course, only generated through Eq.~(\ref{deltaN}), and are given by
\begin{eqnarray}
n_k{(1^+)} = T^{-1}_k(1) - 1 \equiv \pi^2{\rm Ci}^2(-x_1)\;,
\end{eqnarray} 
where we have defined the function
\begin{eqnarray}\label{Cidef}
&& {\rm Ci}(x_j)= {\rm Ai}(-x_j^2){\rm Ai'}(-x_j^2)+{\rm Bi}(-x_j^2){\rm Bi'}(-x_j^2)\,, \\
&& x_j \equiv \frac{j^{1/3}k}{Mq^{1/3}a_j}\,.
\end{eqnarray}
Again, we used $x_j$ in light of Eq.~(\ref{schrodinger}), as the natural argument of the expression of the transmission probability~(\ref{Transmision}). Note that here, the species are only distinguished through the resonance parameters in Eq. (\ref{qfactor}). Normalizing the scale factor at the first zero crossing as $a_1 = 1$, then we can simply write the evolution of the scale factor as $a_j = j^{2/3}$. Thus, the behaviour of $x_j$ with the number of zero crossings goes as $\propto j^{-1/3}$. Then, we can define a typical momentum of the problem, $k_*{(j)}$, related in a very simple way to the resonance parameters $q_Z$, $q_W$ (\ref{qfactor}), as
\begin{equation}\label{typicalMoment}
x_j = \frac{k}{j^{1/3}
k_*{(1)}
}\hV\Rightarrow\hV k_*{(j)} \equiv 
k_*{(1)}j^{1/3}\,,
\end{equation}
where
\begin{eqnarray}\label{typicalInitialMomenta}
k_*{(1)} 
\equiv q^{1/3}M \equiv \left(\frac{2g^2\xi\alpha\kappa\chi_{\rm end}}{4\pi\lambda}\right)^{1\over3}\,M\,,
\end{eqnarray}
with $g = g_2$ and $g_2/\cos\theta_W$ for $W$ and $Z$ bosons, respectively. Since $k_*{(j)}$ is the natural scale for the momenta of the problem, its order of magnitude should coincide simply with the one obtained via the Heisenberg uncertainty principle, see Eq.~(\ref{timea}), as it is indeed the case since
\begin{eqnarray}
k{(j)} \sim a_j(\Delta t_a)^{-1} \equiv \frac{j^{\frac{1}{3}} k_*{(1)}}{2^{1/3}} \approx k_*{(j)}\,.
\end{eqnarray}
Notice that the typical momenta range will be red-shifted because of the expansion of the Universe and, even the comoving typical moment $k_*$, is not a static quantity but rather depends on $j$.

Let us obtain now the total number density of created particles. Just after the $j$-th scattering, this will be given by
\begin{eqnarray}\label{numberDensity}
n(j^+ \geq 2) &=& \frac{1}{2\pi^2a_j^3}e^{\left\lbrace-\gamma\sum_{i=2}^{j-1}F(i)\right\rbrace}\int dk k^2 n_k(1^+)e^{\left\lbrace 2\pi\sum_{i=1}^{j}\mu_k{(i)}\right\rbrace}\nn
&=& M^3\frac{
k_*^3(1)
}{2a_j^3}e^{\left\lbrace-\gamma\sum_{i=1}^{j-1}F(i)\right\rbrace}\int du u^2{\rm Ci}^2(-u^2)\prod_{l=2}^{j}(2\pi^2{\rm Ci}^2(-u^2/l^{2/3})+1)\,,
\end{eqnarray}
where $k_*(1)^3 \propto g^2$~(\ref{typicalInitialMomenta}) should be evaluated with $g=g_2$ or $g_2/\cos\theta_W$, and $\gamma = \gamma_W$ or $\gamma_Z$, respectively for $W$ or $Z$ bosons. This formula encodes the usual resonant behaviour discovered in the 90's, see Refs.~\cite{KLS94,KLS97}, in which it was implicitely assumed that the produced bosons didn't decay between succesive inflaton zero-crossings. However, as we saw in section~\ref{parametricreheating}, the bosons produced each time the Higgs condensate crosses zero, significantly decay before the next scattering. Therefore, we had to correct our formulas for this effect. Fortunately, this was easily done, since the resonant growth occurs in a step-like form, instantaneously (within a time $\Delta t_a$, see Eq.~(\ref{timea})) when the Higgs condensate crosses around zero, while the perturbative decay of the produced vector bosons occurs during the time just between two successive Higgs zero-crossings. Thus, the occupation number just before the $(j+1)$-th scattering, in terms of the occupation number just after the $j$-th scattering, has been corrected by the factor $\exp\lbrace-\gamma\sum_{i=1}^{j-1}F(i)\rbrace$, which accounts for the accummulated effect of the perturbative decays up to the $j$-th scattering.

The combined effect of the non-perturbative parametric resonant at the non-adiabatic regions at the bottom of the potential, together with the perturbative decay along the adiabatic zone during the rest of the semi-oscillation, give rise to a new phenemenology, as we will inmediately see. Therefore, to emphasize the difference from the usual parametric resonance or instant preheating-like mechanisms, we will call these effect $Combined$ $Preheating$.  Expanding the combination of Airy functions (\ref{Cidef}), it is possible to write
\begin{eqnarray}\label{CiExpansion}
{\rm Ci}^2(x_j) \approx Ce^{-Du^2/j^{2/3}},\hV{\rm and}\hV (2\pi^2 {\rm Ci}^2(-u^2/l^{2/3}) + 1) \approx Ae^{-Bu^2/l^{2/3}}\,,
\end{eqnarray}
where $u \equiv j^{1/3}x_j$ and
\begin{eqnarray}\label{CiCoeff}
C = \frac{(2/3)^2}{\Gamma^2(1/3)\Gamma^2(2/3)},\hV D = \frac{12}{3^{2/3}}\frac{\Gamma(2/3)}{\Gamma(1/3)}\,\hV A = 2 + 2\pi^2C,\hV B = 1 - \frac{(16/3)2\pi^2}{3^{2/3}\Gamma^3(1/3)\Gamma(2/3)}\frac{1}{A}\,.
\end{eqnarray} 
Substituying Eqs.~(\ref{CiExpansion}) and (\ref{CiCoeff}) in Eq.~(\ref{numberDensity}), then we obtain
\begin{eqnarray}\label{occupCombPre}
n(j^+ \geq 2) &\approx& M^3 \frac{e^{-\gamma F_\Sigma(j-1)}k_*^3{(1)}}{2j^2}A^{j-1}C\int du u^2e^{-Du^2}e^{-B(\sum_{i=2}^{j}i^{-2/3})u^2} \nn
&=& M^3 \frac{e^{-\gamma F_\Sigma(j-1)}k_*^3{(1)}}{2j^2}A^{j-1}C\frac{\sqrt{\pi}}{4}\left(D+B\sum_{i=2}^{j}i^{-2/3}\right)^{-3/2}\,.
\end{eqnarray} 
where we have used $a_j = j^{2/3}$, performed the resulting gaussian integral and defined
\begin{eqnarray}
F_{_{\Sigma}}(j) \equiv \sum_{i=1}^{j} F(i)\,,
\end{eqnarray}
for simplicity. Notice that the resonant behaviour is now encoded in the factor $A^{j-1}$, which, for sufficiently great $j$, will finally overtake the decaying factor $e^{-\gamma F_\Sigma(j-1)}$, since $A > 2$. Taking also into account the factor $1/j^2$ due to the expansion of the Universe, the first result we can read from here is that only for those values of $j$ for which $(j-1)\log A - 2\log j > \gamma F_{_\Sigma}(j-1)$, the resonant effect will dominate over both the perturbative decay and the expansion rate.
 
\begin{figure}
\centering
\includegraphics[width=7cm, height=5cm]{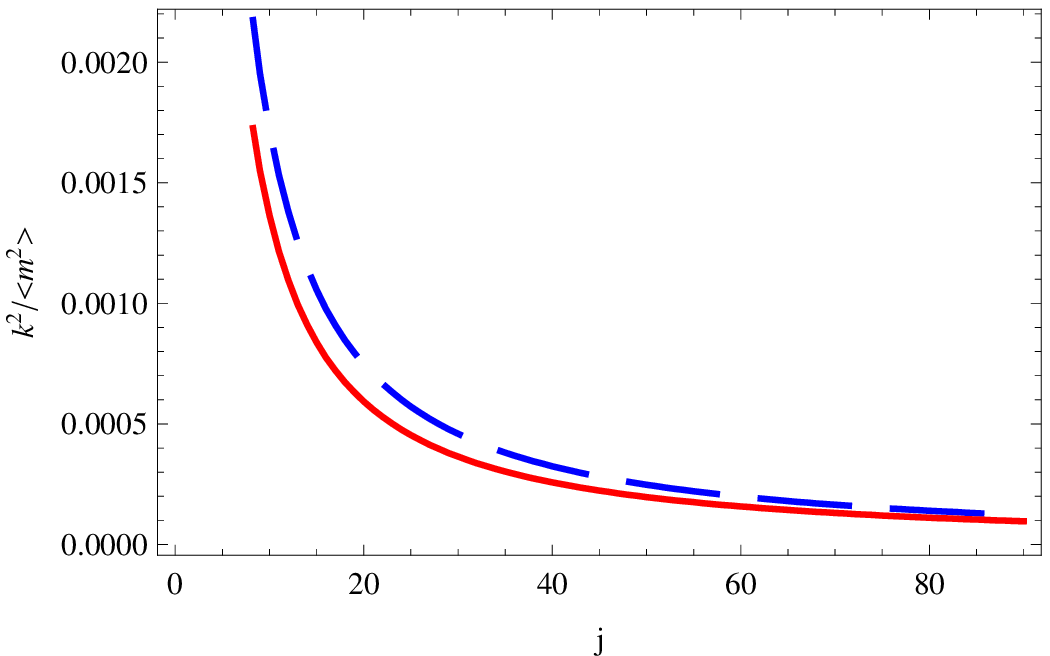}
\hspace{2cm}
\includegraphics[width=7cm, height=5.1cm]{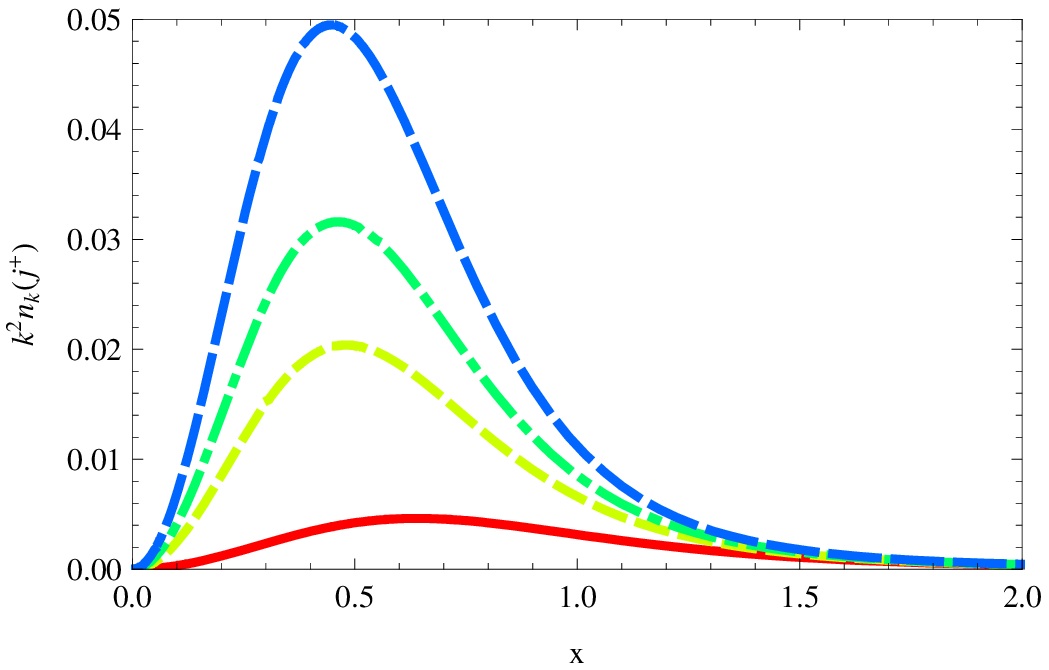}
\caption{Left: The ratio $k^2/\langle m^2 \rangle$ between the typical momenta produced around zero and the average mass in every oscillation for the $W$ (dashed blue line) and $Z$ bosons (continuos red line) as a function of the number of oscillations. This ratio is significantly smaller than 1 for all crossings, which allows us to consider the produced gauge bosons as non-relativisitic. Right: Succesive spectral distributions $k^2n_k(1^+)e^{2\pi\sum_{k=2}^{j}\mu_k(j)}$, at different $j$'s, including the volume factor $k^2$. One can see the predicted~(\ref{pico}) slow displacement of the maxima of the distribution. The x-axis is given in terms of $x = k/(
k_*{(1)})$}
 \label{massmomentaratio}
\end{figure}

Note that inside the integral~(\ref{occupCombPre}), the function $u^2e^{-(D+B\sum_{i=2}^{j}i^{-2/3})u^2}$ has a maximum at a value $u_{p} \equiv (D+B\sum_{i=2}^{j}i^{-2/3})^{-1/2}$, which implies that that the typical (comoving) excited momentum is
\begin{eqnarray}\label{pico}
k_p \approx \frac{
k_*{(1)}}{(D+B\sum_{i=2}^{j}i^{-2/3})^{1/2}}\,.
\end{eqnarray}
In the right hand side of Figure~\ref{massmomentaratio}, one can easily observe this behaviour:  the value of the momentum at which the distribution peaks, slightly moves to smaller values, according to $k_p \propto (D+B\sum_{i=2}^{j}i^{-2/3})^{-1/2}$. Thus, it is noticeable that the typical momentum $k$ of the resonant fluctuations is always of order $k_*{(1)}$, independently of how many oscillations the Higgs performs. The reason of this is that the parametric resonance effect builds up initially from the spectral distribution $n_k(1^+)$, which only filters $k \lesssim k_*{(1)}$, see Eq.~(\ref{formulaCrecimiento}) and Figure~\ref{massmomentaratio}.

On the other hand, the ratio $k_p^2/\langle m \rangle^2_j$ for both $W$ and $Z$ bosons, between the typical momenta produced around zero and the average masses in every oscillation is shown in Fig.~\ref{massmomentaratio}. In particular, it is easy to estimate the evolution in time of such a ratio, in terms of the resonant parameters~(\ref{qfactor}), as
\begin{eqnarray}\label{gaugeBosonsMomenta}
\frac{(k_p/a_j)^2}{\langle m \rangle_j^2} = \frac{
q^{\frac{2}{3}}}{(D+B\sum_{k=2}^{j}k^{-2/3})g^2\left(\frac{3}{2\lambda}\right)\xi F(j)^2} \propto  \frac{1}{g^{2/3}}\frac{1}{j^{1/3}(D+B\sum_{k=2}^{j}k^{-2/3})} \ll 1\,\,\forall j\,,
\end{eqnarray}
Taking into account that the previous ratio (\ref{gaugeBosonsMomenta}) is a decreasing function with $j$, as well as its dependence with the gauge couplings $g^{-2/3}$,  we can conclude that the vector bosons produced at the bottom of the potential are always non-relativistic. This justifies \textit{a posteriori} the calculation of the energy of the fermions as $E_F{(j)} \approx \frac{1}{2}\langle m_{Z,W} \rangle_j$, see Eq.~(\ref{fermionsEnergy}) in section~\ref{parametricreheating}. Extrapolating Eq. (\ref{gaugeBosonsMomenta}) to the case of fermions we realize that the produced particles at each zero crossing would be mainly relativistic, due to the smallness of the Yukawa couplings, being the only exception the production of top quarks. 

The energy density transferred to the fermions between the $j$-th and the $(j+1)$-th scatterings, will be
\begin{eqnarray}\label{deltaRhofer}
\Delta\rho_F{(j)} &=& 6\left[(1-e^{-\gamma_Z F(j)})n_Z(j^+)E_{F_{\rm Z}}(j) + 2(1-e^{-\gamma_WF(j)})n_W(j^+)E_{F_{\rm W}}(j)\right] \nn
&=& \frac{\tilde\epsilon}{2\pi^2}M^2\chi_{\rm end}^2A^{j-1}C\frac{\sqrt{\pi}}{4}\frac{k_*^3}{j^2}\left(D+B\sum_{l=2}^{j}l^{-2/3}\right)^{-\frac32}F(j)\times\nn
&& \times\left((1-e^{-\gamma_ZF(j)})e^{-\gamma_Z\Gamma_\Sigma(j-1)} + 2\cos\theta_W^3(1-e^{-\gamma_WF(j)})e^{-\gamma_W\Gamma_\Sigma(j-1)}\right)\,,
\end{eqnarray}
where we have used the energy of the fermions~(\ref{fermionsEnergy}) and defined a momentum scale independent of the gauge couplings, common to both bosonic species, as
\begin{equation}
k_*\equiv \frac{k_*(1)}{g^{2/3}}\,. 
\end{equation}
The gauge coupling dependence is indeed incorporated on the definition of the parameter
\begin{eqnarray}\label{barEpsilon}
\tilde\epsilon \equiv \frac{3g_2^3
\lambda^{1/2}\pi^2}{(\cos\theta_W)^3\xi^{5/2}(\alpha\kappa\chi_{\rm end})^2}\,,
\end{eqnarray} 
which modulates again the strengh of the effect as $\tilde\epsilon \propto g_2^3$.

The total energy density transferred into the fermions will be
\begin{eqnarray}
\rho_F{(j)} = \sum_{i=1}^{j} \Delta\rho_F{(i)}\left(\frac{i}{j}\right)^{8/3}\,,
\end{eqnarray} 
and the ratio of such an energy to that of the inflaton,
\begin{eqnarray}\label{fracFerm}
\varepsilon_F{(j)} \equiv \frac{\rho_F{(j)}}{\rho_\chi} = \frac{2\pi^2\Big(j+{1\over2}\Big)^2}{M^2\chi_{\rm end}^2}\sum_{i=1}^{j} \Delta\rho_F{(i)}\left(\frac{i}{j}\right)^{8/3} \,,
\end{eqnarray}
with $\Delta\rho_F{(i)}$ given by Eq.~(\ref{deltaRhofer}).
Here we can clearly see the two competing effects; that of the perturbative decay of the bosons, given by the factors of the form $(1-e^{-\gamma F(j)})e^{-\gamma F_{_{\Sigma}}{(j-1)}}$, which tend to decrease the rate of production of bosons and fermions, while the factors $e^{2\pi\mu_k}$ encoded in the form of the gaussian approximation, describe the resonant effect due to the accumulation of previously produced bosons and fermions. Initially, the perturbative decay will prevent the resonance to be effective. However, after a certain number of oscillations (a number that we will estimate next), the resonant effect will overtake the perturbative decays and parametric resonance will be developed as usual, as if the produced bosons would not decay perturbatively during each semi-oscillation.

In order to estimate the time in which the perturbative decays stop blocking the parametric resonance effect, we can evaluate numerically when the expression $e^{-\gamma\Gamma_{\Sigma}(j-1)}$ becomes subdominant versus $e^{2\pi\sum_{i=2}^{j}\mu_k(i)}$. In particular, we can evaluate the ratios, for either $W$ or $Z$
\begin{eqnarray}
\sigma \equiv \frac{2\pi\sum_{i=2}^{j}\mu_k(i)}{\gamma\Gamma_{\Sigma}(j-1)}\,,
\end{eqnarray}
for the fastest growing mode $k = k_p$~(\ref{pico}), and find the number of semi-oscillations $j_{R}$ for which the previous ratio becomes greater than one, $\sigma \geqslant 1$. We find $j_{R} \approx 62$ for the $W$ bosons and $j_R \approx 360$ for the $Z$ bosons. The fact that parametric resonance becomes important much earlier for $W$'s than for $Z$'s is not a surprise, since 
their decay rate~(\ref{widthWE}) differ in a factor $\gamma_Z/\gamma_W \approx 2.4$, which simply means that there are many more $W$ bosons surviving per semioscillation than $Z$ bosons. Therefore, the \textit{combined preheating} of the $W$ bosons is much faster driven into the parametric-like behaviour, while the evolution of the $Z$ bosons is much more affected by the perturbative decays, delaying (or even completely preventing) the development of parametric resonance. Obviously, after a dozen of oscillations, the transfer of energy from the inflaton to the gauge bosons will be completely dominated by the channel into the $W$ bosons, since by that moment they will be fully resonant while the $Z$ bosons will still severely affected by their perturbative decay. 

Finally, to conclude this section and achieve an overall complete picture of all the details, let us also estimate the transfer of energy from the inflaton to the gauge bosons. In particular, the total energy transferred to them just after the $j$-th scattering, $\rho_B{(j)}$, is given simply by 
\begin{equation}
\rho_B(j) = 3\left(n_Z(j^+)\langle m_Z \rangle_{j} + 2n_W(j^+)\langle m_W \rangle_{j}\right)\,,
\end{equation}
where we have used the fact that the gauge bosons are non-relativistic and have 3 polarizations. Therefore, the ratio of the energy of the gauge bosons to the energy of the inflaton, can be expressed as
\begin{equation}\label{fracBosons}
\varepsilon_B{(j)} \equiv \frac{\rho_B{(j)}}{\rho_\chi} =  \Big(j+\frac{1}{2}\Big)^2\left(\frac{1
}{\cos\theta_W}\right)^2\frac{\sqrt{\pi}\,\tilde\epsilon k_*^3 F(j)\,A^{j-1}C}{4j^{2}\left(D+B\sum_{i=2}^{j}i^{-2/3}\right)^{3/2}}\left(e^{-\gamma_ZF_\Sigma(j)} + 2\cos\theta_W^3e^{-\gamma_WF_\Sigma(j)}\right)\,,
\end{equation}
where we have used Eq.~(\ref{occupCombPre}) and $\tilde\epsilon$, defined in Eq.~(\ref{barEpsilon}), modulates again the amplitude of this growing function.

Using Eqs. (\ref{fracFerm}) and (\ref{fracBosons})  , we can estimate the time in which finally the energy of the inflaton would be transferred efficienty to the fermions or the bosons. Defining that moment, respectively, like $\varepsilon_F{(j_{\rm eff})} \equiv 1$ and $\varepsilon_B{(j_{\rm eff})} \equiv 1$, one obtains
the numbers in Table I. Note that the bosons receive the transfer of energy from the inflaton before the fermions, since by the time that parametric resonance overtakes the perturbative decay, the fraction of bosons decaying (per semi-oscilation) into fermions is very small and, therefore, the fraction of newly added fermions is less and less important, while the amount of produced bosons is more and more prominent. Note also that the number of oscillations $j_{\rm eff}$ required for an efficient transfer of energy, depends on the parameter $\lambda$, although the overall order of magnitude does not change appreciably.

\begin{table}\label{ratio}
\begin{center}
\begin{tabular}{||c c c| c c c c c c c c c c c ||}
\hline\hline 
\noalign{\smallskip}
&$\lambda$& && 0.2  && 0.4  && 0.6 && 0.8 && 1.0&  \\[1mm]
\hline 
&$j_{\rm eff}^{(B)}$& && 74 && 64 && 60 && 57 && 55&  \\[1mm]
&$j_{\rm eff}^{(F)}$& && 79 && 69 && 64 && 61 && 59 &  \\[1mm]
\hline\hline
\end{tabular}
\caption{Number of semi-oscillations of the Higgs required, as a function of $\lambda$, for an efficient transfer of energy from the inflaton to the gauge fields and/or to the fermions.}
\end{center}
\end{table}

Unfortunately, as we will see in the next subsection, before reaching the stage in which $\epsilon_{F,B} \sim 1$, the backreaction of the produced gauge fields into the homogeneous Higgs condensate will become significant, and it will have to be taken into account.

\section{Backreaction}\label{backreaction}

Let's now calculate the backreaction from the $W$ and $Z$ bosons into the Higgs condensate. Neglecting the vectorial nature of the bosons, the effective equation of the Higgs condensate can be written, in the Hartree approximation, as
\begin{eqnarray}
\ddot\chi +3H\dot\chi -\frac{1}{a^3}\nabla^2\chi + \left[M^2 + \frac{g^2M_p^2}{4\xi}\frac{1}{\chi}\frac{\partial}{\partial\chi}(1-e^{-\alpha\kappa|\chi|})\langle\varphi^2\rangle\right]\chi = 0\,,
\end{eqnarray}
where there is a $\varphi$ field for each polarization of each gauge boson species, such that $g^2 = g_2^2$ and $g^2 = g^2_2/\cos^2\theta_W$, as usual, for the $W$ and $Z$ bosons, respectively. From here, performing the derivative, one obtains for the effective Higgs frequency
\begin{eqnarray}
\omega^2 = M^2 + \frac{\alpha g^2 M_p}{4\xi|\chi|}e^{-\alpha\kappa|\chi|}\langle\varphi^2\rangle\,.
\end{eqnarray}
where the second term in the $r.h.s.$ should be summed over polarizations and species. For the fraction of time of each semi-oscillation, during which the Higgs frequency evolves adiabatically, we can use the correlation function
\begin{eqnarray}
 \langle \varphi_k\varphi_{k'}^*\rangle = (2\pi)^3|\varphi_k|^2\delta(k-k')\,,
\end{eqnarray}
with $\varphi_k(t)$ expressed as 
\begin{eqnarray}
a^{3/2}\varphi_k(t) = \frac{\alpha_k(t)}{\sqrt{2\omega(k)}}e^{-i\int_0^t\omega_kdt'} + \frac{\beta_k(t)}{\sqrt{2\omega(k)}}e^{+i\int_0^t\omega_kdt'}\,.
\end{eqnarray}
Thus, one can compute the expectation value of the bosonic fields (components)
\begin{eqnarray}
\langle \varphi^2\rangle &\equiv& \frac{1}{2\pi^2a^3}\int dk k^2|\varphi_k|^2 = \frac{1}{2\pi^2a^3}\int\frac{dk k^2}{\omega_k}\left(\frac{1}{2}+|\beta_k|^2 + {\rm Re}\lbrace\alpha_k\beta_k^*e^{-i2\int^t\omega dt'+{\rm Arg}\,\alpha_k+{\rm Arg}\,\beta_k}\rbrace\right) \nn
&\approx& \frac{1}{2\pi^2a^3}\frac{2\sqrt\xi}{gM_p}\frac{1}{\sqrt{1-e^{-\alpha\kappa|\chi|}}}\int dk k^2\,n_k\,\Big[1+\cos\Big(\frac{2\pi}{M}\sum_j\langle\omega\rangle_j + {\rm Arg}\,\alpha_k+{\rm Arg}\,\beta_k\Big)\Big]\,,
\end{eqnarray} 
where, to obtain the last exppresion we have used $\omega_k = \frac{gM_p}{2\sqrt\xi}\sqrt{1-e^{-\alpha\kappa|\chi|}}$, $|\beta_k|^2 = |\alpha_k|^2 - 1 = n_k$ and $\int\omega_k(t')dt' = (\pi/M)\sum_{j=1}^{n}\langle\omega\rangle_j$, with $\langle\omega\rangle_j = \frac{M}{\pi}\int_{t_j}^{t_{j+1}} dt'\omega(t')$. Following~\cite{KLS97}, since we don't know the accummulated phases of $\alpha_k$ and $\beta_k$, we will write
\begin{eqnarray}
\langle \varphi^2\rangle \approx \frac{2\sqrt\xi}{gM_p}\frac{n_\varphi}{\sqrt{1-e^{-\alpha\kappa|\chi|}}}\Big[1+A\cos\Big(\frac{2\pi}{M}\sum_j\langle\omega\rangle_j\Big)\Big]\,,
\end{eqnarray} 
with $A<1$ and $n_\varphi = (2\pi^2a^3)^{-1}\int dk k^2 n_k$.

From here, one can define the effective frequency of the Higgs condensate as
\begin{eqnarray}
\omega^2 \equiv M^2 + \frac{\alpha g\,n_\varphi}{2\sqrt\xi|\chi|}\frac{\left[1+A\cos\left(\frac{2\pi}{M}\sum_j\langle\omega\rangle_j\right)\right]}{\sqrt{e^{2\alpha\kappa|\chi|}-e^{\alpha\kappa|\chi|}}}\,.
\end{eqnarray}
The backreaction of the gauge boson fields over the Higgs field, will be non-negligible when the last term in the r.h.s. of the previous expression becomes of the order of $M^2$. In terms of the number densities of the $Z$ and $W$ bosons, i.e. summing the contribution over polarizations and species of all the fields that back react, this will happen at a time $t_j = j\pi/M$,
\begin{eqnarray}
{\rm Backreaction}\hV \Leftrightarrow\hV \Big(n_Z(j)/\cos\theta_W + 2n_W{(j)}\Big) \gtrsim \frac{2\sqrt\xi |\chi(t_j)|(\alpha\kappa|\chi(t_j)|)^{1/2}M^2}{3\alpha g_2}\,,
\end{eqnarray}
where we have expanded $\sqrt{e^{2\alpha\kappa|\chi|}-e^{\alpha\kappa|\chi|}} \approx (\alpha\kappa|\chi|)^{1/2}$, which is certainly accurated after a couple of dozens of ocillation, since $|\chi(t_n)| \propto \frac{1}{j}$. Substituying the averaged value per semi-oscillation $\chi(t) \rightarrow \langle \chi(t) \rangle_j$, then we will take $\alpha\kappa|\chi(t_j)| \rightarrow \frac{\alpha\kappa\chi_{\rm end}}{\pi j}(\frac{1}{\pi}\int_{0}^{\pi}\sin(x)) = \frac{2}{\pi}\frac{\alpha\kappa\chi_{\rm end}}{\pi j}$. Using the analytical expressions for the occupation numbers (\ref{occupCombPre}) we can translate the above condition into the following one
\begin{eqnarray}
\left(e^{-\gamma_Z\Gamma_\Sigma(j-1)}/\cos^3\theta_W+2e^{-\gamma_W\Gamma_\Sigma(j-1)}\right)\frac{A^{(j-1)}}{j^{1/2}\left(D+B\sum_{i=2}^{j}i^{-3/2}\right)} \geq \frac{32\sqrt{6}\xi^{3/2}\log(1+2/\sqrt{3})^{3/2}}{3\lambda^{1/2} \alpha^2 g_2^3\pi^{7/2}C\,k_*^3}\,.
\end{eqnarray} 
Thus, if we find numerically the number of semi-oscillations of the Higgs, $j_{\rm backr}$, for which $j>j_{\rm backr}$ fullfills the above condition, then we know the moment in which backreaction of the bosonics fields becomes significant, $t_{\rm backr} \approx \pi j_{\rm backr}/M$. Note that the above condition depends on $\lambda$ both in the left- and right-hand sides. In particular, the $\lambda$ dependence is rather weak in the constant of the right-hand side of the inequality, since it goes as $\lambda^{1/4}$, while in the left-hand side, it enters through the exponentials so it can change the number $n_{\rm backr}$ in a more significant manner. Taking values of $\lambda$ between 0.2 and 1.0, we 
obtain the numbers in Table II.

\begin{table}
\begin{center}
\begin{tabular}{| | c c c | c c c c c c c c c c||}
\hline\hline 
\noalign{\smallskip}
&$\lambda$& & 0.2 & & 0.4 & & 0.6 & & 0.8 & & 1.0 & \\[1mm]
\hline
&$j_{\rm backr}$& & 67 & &57 & &52 & &50 & & 48& \\[1mm]
\hline\hline
\end{tabular}
\caption{Number of semi-oscillations of the Higgs required, as a function of $\lambda$, for the backreaction of the gauge fields into the Higgs background to become significant.}
\end{center}
\end{table}

We clearly see that backreaction seems to become important at a time slightly earlier than that at which we were expecting the Higgs to have transferred efficiently its energy to the bosons and fermions. This means that our analytical estimates of these transfers were biased, and a careful numerical study of the process is required. Beyond backreaction, the strength of the resonance very quickly decreases due to the increased frequency of oscillations of the Higgs. Eventually, the broad resonance driving the production of gauge bosons and thus their decay into SM particles becomes a narrow resonance and finally shuts off. From then on, the inflaton will oscillate like a matter field while the produced particle will redshift as radiation, its effect on the expansion becoming negligible after a few hundred oscillations.

\section{Conclusions}\label{conclusions}

We have studied the different stages of reheating after inflation in a model where the role of the inflaton is played by the Higgs field of the Standard Model of particle physics with a non-minimal coupling to gravity. Inflation in this model takes place at the GUT scale, along the lines of the Starobinsky model of inflation since a conformal transformation makes these two models indistinguishable from the point of view of inflation. The usual difficulty with large self couplings of the Higgs is tamed here by the inclusion of a large non-minimal coupling to gravity, $\xi\sim10^5$, which nevertheless does not leave any signature at low (electroweak) scales due to the fact that the Higgs field acquires a vacuum expectation value and does not evolve at present, while the local spacetime curvature is negligible.

The advantage of this model of inflation for the study or reheating after inflation is that all the couplings of the Higgs-Inflaton to matter fields are known at the electroweak scale, and can be extrapolated to the GUT scale using the renormalization group equations, and therefore one can study in detail the process of reheating of the Universe, without having to impose ad hoc assumptions about their values. The surprise is that the process becomes more complicated than expected, and a series of subsequent stages take place, where essentially all different types of particle production mechanisms at preheating occur. Moreover, since the Standard Model couplings of the Higgs to gauge and matter fields are non-negligible, nor are their couplings among themselves, the process of non-perturbative decay via parametric resonance is mixed with the usual perturbative decays of the decay products, which complicates things significantly.

Inflation ends at values of the Higgs field of order the Planck scale and goes through a brief stage of tachyonic preheating soon after the end of inflation. The passage is so short that particle production is not significant at that stage. The same occurs with the production at the inflection point. Finally the Higgs-Inflaton field starts oscillating around the minimum of its potential with a curvature scale of order $10^{13}$ GeV. At this stage, particle production occurs whenever the Higgs passes through zero, creating mostly vector gauge bosons $W$ and $Z$. These gauge bosons acquire a large mass while the Higgs increases towards maximum amplitude and start to decay into all Standard Model leptons and quarks within half a Higgs oscillation, rapidly depleting the occupation numbers of gauge bosons, like in instant preheating. However, the fraction of energy of the Higgs that goes into SM particles is still very small compared with the energy in the oscillations, and therefore the non-perturbative decay is slow. This implies that a relatively large number of oscillations take place before a significant amount of energy is transferred to the gauge bosons and fermions. 

The amplitude of Higgs boson oscillations decreases as the Universe expands in a matter-like dominated stage with zero pressure. Eventually, this amplitude is small enough that the gauge boson masses are not large enough for inducing a quick decay of the gauge bosons and these start to build up their occupation numbers very rapidly via parametric amplification. The question whether this effect can give rise to the production of a significant Gravitational Wave Background (GWB) potentially observable today remains to be addressed. Several papers have studied recently such an issue in the chaotic and hybrid models of inflation~\cite{GarciaBellido:2007dg}, but in the present model, we don't have simply a parametric resonance phenomena but a combined preheating effect which, perhaps, could modify the properties of such a GWB. Similar arguments would affect also the production of magnetic fields at preheating~\cite{DiazGil:2007dy} or even electroweak baryogenesis~\cite{EWB}. 

After about a hundred oscillations the gauge bosons produced backreact on the Higgs field and the resonant production of particles stops. The Higgs field acquires a large mass via its interaction with the gauge condensate and preheating ends. From there on, both Higgses and gauge fields decay perturbatively until their energy is transferred to SM particles. Since the stage after backreaction is very non-linear and non-perturbative, it cannot be solved analytically and we have to resort to numerical studies in the lattice. We leave the description of our numerical studies to a future publication.

\

{\bf Note added}: Upon completion of this paper, we received through the arXiv the preprint of Bezrukov et al.~\cite{Bezrukov:2008}, where they also study preheating in the $\nu$MSM. Although the formalism is common to both, our conclusions are somewhat different from those of Ref.~\cite{Bezrukov:2008}. We find that the Higgs decay into gauge bosons is significantly faster, and that backreaction occurs much before thermalization. We thus think it is not possible to determine the reheating temperature without a careful numerical analysis with lattice simulations.

A few days after this work was completed, Ref.~\cite{DeSimone:2008} was also posted in the arXiv, performing an analysis of the 2-loop quantum corrections to the running of all the parameters involved in the model. This paper allows for a different range of the Higgs self-coupling, which is compatible with the range~(\ref{HiggsRange}) although more restricted. The main result of Ref.~\cite{DeSimone:2008} is a relationship between the Higgs mass and the spectral index which, in principle, could be tested in the future against data from PLANCK and the LHC. Similar conclusions where found in Ref.~\cite{Magnin:2008}. 

\subsection*{Acknowledgements}

We would like to thank Geneva University, SISSA-Trieste and MPI-Munich for hospitality during the development of parts of this research. DGF is supported by a FPU contract with Ref. AP2005-1092 and JR by an I3P contract. We also acknowledge financial support from the Madrid Regional Government (CAM) under the program HEPHACOS P-ESP-00346, and the Spanish Research Ministry (MEC) under contract FPA2006-05807. The authors participate in the Consolider-Ingenio 2010 CPAN (CSD2007-00042) and PAU (CSD2007-00060), as well as in the European Union 6th Framework Marie Curie Network ``UniverseNet" under contract MRTN-CT-2006-035863.


\end{document}